      [1999/12/01 v1.4c Il Nuovo Cimento]

\documentclass[twocolumn,varenna]{cimento}

\usepackage{graphicx}  
\usepackage{textcomp}
\usepackage[square,sort&compress,comma,numbers]{natbib}
\title{Colloidal Shape Effects in Evaporating Drops}

\author{P.~J.~Yunker$^{1}$,T.~Still$^{1,2}$,A.~G.~Yodh$^{1}$}

\institute{$^{1}$Department of Physics and Astronomy, University of Pennsylvania, Philadelphia PA 19104, USA}
\institute{$^{2}$Complex Assemblies of Soft Matter, CNRS-Rhodia-University of Pennsylvania UMI 3254, Bristol, Pennsylvania 19007, USA}

\shortauthor{P.~J.~Yunker,T.~Still,A.~G.~Yodh}

\PACSes{47.57.J-, 46.70.Hg, 47.55.nb, 47.55.np}
\begin{document}

\maketitle

\begin{abstract}
We explore the influence of particle shape on the behavior of evaporating drops. A first set of experiments discovered that particle shape modifies particle deposition after drying. For sessile drops, spheres are deposited in a ring-like stain, while ellipsoids are deposited uniformly. Experiments elucidate the kinetics of ellipsoids and spheres at the drop's edge. A second set of experiments examined evaporating drops confined between glass plates. In this case, colloidal particles coat the ribbon-like air-water interface, forming colloidal monolayer membranes (CMMs). As particle anisotropy increases, CMM bending rigidity was found to increase, which in turn introduces a new mechanism that produces a uniform deposition of ellipsoids and a heterogeneous deposition of spheres after drying. A final set of experiments investigates the effect of surfactants in evaporating drops. The radially outward flow that pushes particles to the drop's edge also pushes surfactants to the drop's edge, which leads to a radially inward flow on the drop surface. The presence of radially outward flows in the bulk fluid and radially inward flows at the drop surface creates a Marangoni eddy, among other effects, which also modifies deposition after drying.
\end{abstract}

\maketitle
\section{Introduction}

\begin{figure}
\resizebox{\columnwidth}{!}{\includegraphics{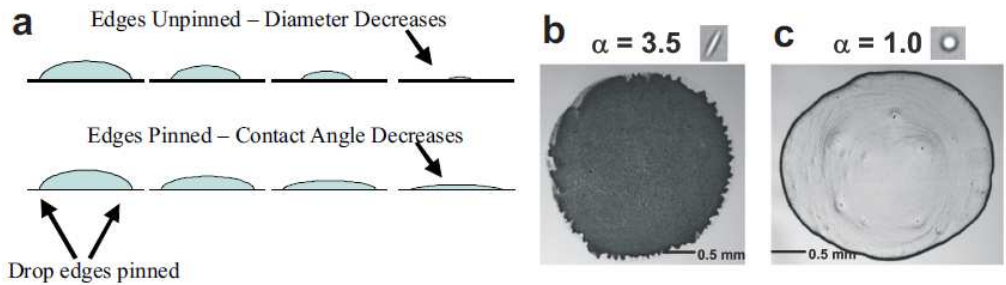}}
\caption{\label{droppin} a. Cartoon depicting evaporating drops with edges unpinned (top) and pinned (bottom). b. Image of the final distribution of ellipsoids after evaporation. c. Image of the final distribution of spheres after evaporation. Images of a single ellipsoid and a single sphere are shown above (b) and (c).}
\end{figure}

In this contribution we describe experimental variations on the so-called coffee-ring effect. If you have spilled a drop of coffee and left it to dry, then you might have observed a ring-shaped stain. Specifically, the stain is darker near the drop edges compared to the middle (Fig.\ \ref{droppin}c).  This phenomenon is the coffee-ring effect; it is produced by the interplay of fluid dynamics, surface tension, evaporation, diffusion, capillarity, and more. Briefly, as a drop evaporates, its edges easily become pinned and cannot recede towards the middle of a drop, i.e., the diameter of a pinned drop does not decrease (Fig.\ \ref{droppin}a). This effect is perhaps surprising considering that fluid regions near the edges of a drop are thinner than in the middle. Thus, fluid flows from the middle of the drop to the edge of the drop to replenish evaporated water. This flow readily carries suspended particles, moving them from the middle of the drop to its edges, thus producing a coffee-ring.

Why care about the coffee-ring effect? A drop of evaporating water is a complex, difficult-to-control, non-equilibrium system. Along with capillary flow, the evaporating drop features a spherical-cap-shaped air-water interface and Marangoni flows induced by small temperature differences between the top of the drop and the contact line \cite{contactline_PRE,suspendedparticlespreading}. Thus, to understand the coffee-ring effect, one must understand pinning effects, fluid dynamics, particle-substrate interactions, substrate-fluid interactions, and more. Indeed, intellectual challenges have motivated us to understand this complex, far-from-equilibrium system, and the effects of each of these parameters.

Of course, if the coffee-ring effect were only present in coffee and tea, its practical importance would be minimal. In fact, the coffee ring effect is manifest in systems with diverse constituents ranging from large colloids \cite{coffeering_nagel_nat,contactline_PRE,drop_patterns_PRE,inside_coffee_ring} to nanoparticles \cite{monolayer_jaeger} to individual molecules (e.g., salt) \cite{evaporation_salt}. Due to its ubiquity, the coffee-ring manages to cause problems in a wide range of practical applications which call for uniform coatings, such as printing \cite{printing_convec_assemb}, genotyping \cite{DNA_graft_evap,evap_bio}, and complex assembly \cite{printing_polymers}. Paint is another system susceptible to the coffee-ring effect. To avoid uneven coatings, paints often contain a mixture of two different solvents. One is water, which evaporates quickly and leaves the pigment carrying particles in a second, thicker solvent. The particles are unable to rearrange in this viscous solvent and are then deposited uniformly. Unfortunately, this second solvent also evaporates relatively slowly (one reason why it might be boring to watch paint dry). While a number of schemes to avoid the coffee-ring effect have been discovered \cite{monolayer_jaeger,marangoni_reverses_coffeering,coffeering_reverse_capforce,surfactant_rev_CReffect,marangoni_benard_PRL,disjoiningpressure_assemb}, these approaches typically involve significant modifications of the system. Thus, the discovery of relatively simple ways to avoid the coffee-ring effect and control particle deposition during evaporation could greatly benefit a wide range of applications.

To this end, we asked (and answered) a question: does particle \textit{shape} affect particle deposition \cite{coffee_ring_ellipsoids}? At first glance, it may appear that shape should not matter. Colloidal particles of any shape are susceptible to the radially outward flow of fluid that drives the coffee-ring effect. However, changing particle shape dramatically changes the behavior of particles on the air-water interface. In fact, smooth anisotropic ellipsoids deform the air-water interface while smooth isotropic spheres do not \cite{ellipsoid_wetting_yodh,capillary_interactions_ellipsoids_yodh,interface_attract_plates,interface_anisotropicparticles,ellipsoids_interface_rheology,emulsions_ellipsoids_softmatter,interface_att_furst}. Deforming the air-water interface, in turn, induces a strong interparticle capillary attraction between ellipsoids. This capillary attraction causes ellipsoids to form a loosely-packed network that can cover the entire air-water interface, leaving ellipsoids much more uniformly distributed when evaporation finishes (Fig.\ \ref{droppin} b). Conversely, spheres pack densely at the drop's edge, producing a coffee-ring when evaporation has finished (Fig.\ \ref{droppin} c). Thus, particle \textit{shape} can produce uniform coatings.

The remainder of this review is organized as follows. First, we discuss the different interfacial properties of spheres and ellipsoids, as well as the methods to make anisotropic particles. Then, we discuss our investigation of particle behavior in evaporating sessile drops and the coffee-ring effect. Much of this work is described in a recent publication \cite{coffee_ring_ellipsoids}. In particular, we demonstrate that particle shape strongly affects the deposition of particles during evaporation. Next, we investigate the role of particle shape in evaporating drops in confined geometries, and we show how to extract the bending rigidity of the membranes formed by particles adsorbed on the air-water interface. Much of this work is described in another publication \cite{confined_evap_ellipsoids}. Finally, we shift focus to discuss the effects of surfactants on evaporating colloidal drops. We show that surfactants lead to a radially inward flow on the drop surface, which creates a Marangoni eddy, among other effects, which leads to differences in drying dynamics. Some of this work was published recently \cite{marangoni_eddies}.  As a whole, this review attempts to present these experiments in a unified fashion.

\section{Anisotropic Particles}

\subsection{Capillary Interactions - The Young Laplace Equation}

At small packing fractions, i.e., outside the range which would lead to formation of crystalline or liquid crystalline phases, the diffusion and hydrodynamics of spheres and ellipsoids are only modestly different \cite{ellipsoid_brownianmotion}. Further, both spheres and ellipsoids will adsorb onto the air-water interface; the binding energy of micron-sized particles to the air-water interface depends primarily on the interfacial area covered by the particle and the contact angle, quantities which are similar for spheres and ellipsoids.  The binding energy for a micron-sized particle is $\sim10^{7}k_{B}T$, where $k_{B}$ is the Boltzmann constant and $T$ is temperature \cite{adsorption_review}. Once adsorbed onto the air-water interface, however, the behaviors of spheres and ellipsoids are dramatically different \cite{free_energy_sessile,cap_int_emulsions}. Anisotropic particles deform interfaces significantly, which in turn produces very strong interparticle capillary interactions  \cite{ellipsoid_wetting_yodh,capillary_interactions_ellipsoids_yodh,interface_attract_plates,interface_anisotropicparticles,ellipsoids_interface_rheology,emulsions_ellipsoids_softmatter,interface_att_furst,curvaturedriven_rod_stebe}. These deformations have been predicted \cite{roughness_cap_attract,ellipsoids_furst, cap_int_emulsions, free_energy_sessile, ellipsoid_interface_theory, aniso_capil_sim, ellipsoid_mono_rheo_theory, ellipsoid_fluid_interface,floating_prolate, capil_force_interface, aniso_capil_assem, interface_attract_plates, interface_jamming, interface_capillaryeffects, interface_deformation_nagayama, danov10} and have been experimentally observed via techniques such as ellipsometry and video microscopy \cite{ellipsoid_wetting_yodh,capillary_arrows,cap_def_seed,cylinder_deformations}. Two particles that deform the air-water surface will move along the interface to overlap their deformations and thus minimize total system (particles plus interface) energy. This preference at the interface effectively produces a strong interparticle attraction, which has been measured to be hundreds of thousands times greater than thermal energy for micron size particles \cite{capillary_interactions_ellipsoids_yodh,interface_att_furst}.

The interfacial deformations can be understood from expanded solutions of the Young-Laplace equation \cite{roughness_cap_attract,ellipsoids_furst, cap_int_emulsions, free_energy_sessile, ellipsoid_interface_theory, aniso_capil_sim, ellipsoid_mono_rheo_theory, ellipsoid_fluid_interface,floating_prolate, capil_force_interface, aniso_capil_assem, interface_attract_plates, interface_jamming, interface_capillaryeffects, interface_deformation_nagayama, danov10,sessile_evap_review,cap_attr_review_stebe}. The Young-Laplace equation minimizes the energy associated with a surface, and thus relates the pressure difference across the surface to the curvature of the surface. Specifically, the Young-Laplace equation is a force balance statement:  $\gamma H=p_{air}-p_{water}$, where $\gamma$ is surface tension, $H$ is mean curvature of the interface, $p_{air}$ is the pressure in the air, and $p_{water}$ is the pressure in the water \cite{roughness_cap_attract}. For length scales smaller than the capillary length (i.e., the length scale at which the Laplace pressure from surface tension is equal to the hydrostatic pressure due to gravity, $2$ mm for water), gravitational effects can be ignored, and the pressure drop across the surface is zero, implying that the mean curvature everywhere is zero. The mean curvature can be expressed as $H=\Delta h$, where $\Delta$ is the Laplacian and $h$ is the height of the surface. Thus, $\Delta h = 0$.

\begin{figure}
\resizebox{\columnwidth}{!}{\includegraphics{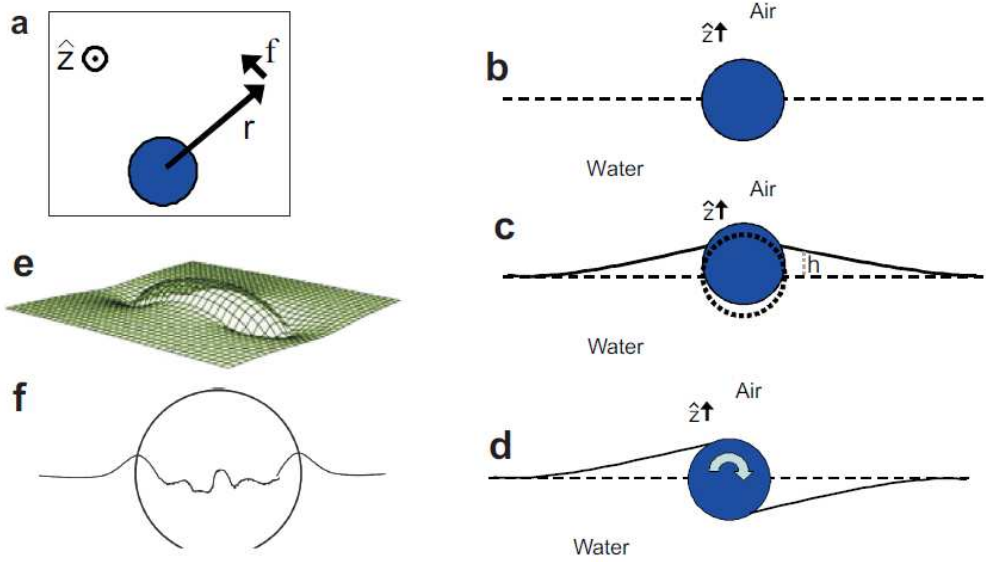}}
\caption{\label{YLeq} a. Overhead cartoon of particle, defining radial position, $r$, polar angle $\phi$, direction $z$ normal to undisturbed interface. b. Side-view cartoon of a smooth sphere on an undeformed interface. c-d. Side-view cartoons of particles on interfaces causing monopolar (c) and dipolar (d) deformations. The interfacial deformation height, $h$, is defined in (c). Dashed lines indicate the position of the undeformed interface and particle position. In (d), the arrow indicates the direction of the torque which produced a dipolar deformation. e. Rendering of a quadrupolar interfacial deformation around an ellipsoid. f. Cartoon of a heterogeneously pinned three-phase contact line on a sphere. This contact-line-roughness deforms the air-water interface with a quadrupolar symmetry, similar to the shape-based deformations characteristic of ellipsoids.}
\end{figure}

When a particle attaches to the air-water interface, boundary conditions are created for the air-water interface. Theoretically, one seeks to solve the Young-Laplace equation with these boundary conditions. In this case, it is useful to first rewrite the Young-Laplace equation in polar coordinates, $\Delta h(r,\phi)=(r^{-1} \partial_{r} r \partial_{r}+r^{-2}\partial_{\phi}^{2})h(r,\phi)=0$ (see Fig.\ \ref{YLeq} a-c). This problem is similar to potential problems in electrostatics and can be solved by separation of variables, i.e., with the ansatz $h(r,\phi)=R(r)\Phi(\phi)$. Substitution for $h(r,\phi)$ leads to $(r^{-1} \partial_{r} r \partial_{r}R(r))\Phi(\phi)+(r^{-2}\partial_{\phi}^{2}\Phi(\phi))R(r))=0$.  Since this equation must hold as $r$ and $\phi$ are varied independently, each term in the equation must equal the same constant, which is leadingly termed $m^{2}$. Thus, $\partial_{\phi}^{2}\Phi(\phi)=m^{2}\Phi(\phi)$ and
$r \partial_{r} r \partial_{r}R(r)=m^{2}R(r)$, with solutions $\Phi=A_{m} cos(m(\phi-B_{m}))$ and $R=C_{m}r^{-m}$, where $A_{m}$, $B_{m}$, and $C_{m}$ are determined by boundary conditions.

The monopole term $m=0$ is only non-zero when the height of the interface near the particle is uniformly lowered (or raised) (Fig.\ \ref{YLeq} c). The monopole term is only stable for the particle in an external field (e.g., gravity); however, for typical colloidal particles the gravitational buoyancy forces are not significant and this term is zero. The dipole term $m=1$ corresponds to a situation wherein the height of the interface is lower on one side of the particle compared to the opposite side (Fig.\ \ref{YLeq} d). Thus, this situation can be quickly relaxed by rotating the particle, i.e., lowering the interface on the high side and raising it on the low side. The dipole term is only stable when an external torque is applied; since no external torques act on the particles, this term also is zero. Therefore, the lowest allowed term is the quadrupole term ($m=2$), i.e., $h(r,\phi) \approx A_{2}cos(2(\phi-B_{2}))C_{2}r^{-2}$ (Fig.\ \ref{YLeq} e).

Notice, this derivation has not mentioned anisotropic boundary conditions. In fact, the quadruploar form for $h(r,\phi)$ is applicable in general to any deformation of the air-water interface (absent external forces and torques) that arises at the particle surface. The air-water interface can be deformed on a sphere, if the three-phase contact line is heterogeneously pinned (see Fig.\ \ref{YLeq} f) \cite{roughness_cap_attract,aniso_janus_o_w_dlee,janus_o_w_dlee,nonsphericalamphiphilic_dlee,janus_bubbles_air_water_dlee,slowadsorption_oninterface}. This effect produces a quadrupolar profile of the interfacial height. However, the linear size of deformation, i.e., $\Delta h$, the maximum value of $h$ minus the minimum value of $h$, from contact-line-roughness is typically much smaller than the linear size of the deformation from shape-based-roughness (for example, see reference \cite{interface_att_furst}). Of course, if one applies Young's conditions for the three-phase contact line on the solid particle, one ``ideally'' obtains a circle contact line on the sphere and a much more complicated shape on an anisotropic particle such as an ellipsoid.  On the ellipsoid, this leads to height variations of $h(r,\phi)$ that are of order the particle size.

The interaction potential between two particles is related to the excess surface area created by these deformations \cite{roughness_cap_attract}. For a single particle ($i$), the deformation energy is $U_{i}=\gamma \delta A_{i}$, where $\delta A_{i}$ is the excess surface area due to interfacial deformation, which is proportional to the deformation size squared, i.e., $\delta A_{i}\propto \Delta h^{2}$ The interaction energy of two particles ($i,j$) is $U_{ij}=\gamma \delta (A_{ij}-A_{i}-A_{j})$, where $A_{ij}$ is the excess surface area due to both particle $i$ and particle $j$ (which is dependent on the particle positions and orientations), and $A_{i}$ and $A_{j}$ are the excess areas due to particle $i$ and $j$ alone. For smooth spheres $A_{ij}=A_{i}+A_{j}$, since the interface is not significantly deformed, so $U_{ij}\approx0$. For ellipsoids (or rough spheres), $U_{ij}\approx -12 U_{i} cos[2(\phi_{i}-\phi_{j})]r^{-4}$, where $\phi_{i}$ and $\phi_{j}$ are the angular orientations of ellipsoids $i$ and $j$. The attractive strength decays as $r^{-4}$ and depends on the coefficient term $U_{i}$, which, in turn, depends on the deformation size squared, i.e., $U_{i}\propto \Delta h^{2}$. Thus, the strength of this attraction ultimately depends strongly on the size of the deformation at the surface of the particle. For example, $1$ micron diameter particles that induce interfacial deformations of $\Delta h = 100$nm and an interparticle separation of $2$ microns will produce an attraction with strength $U_{ij}\approx 2 \times 10^{5} kT$. For micron-sized ellipsoids, the binding energy from capillary attraction is $\sim10^{5}k_{B}T$ \cite{capillary_interactions_ellipsoids_yodh,interface_att_furst}.

To summarize, spheres and ellipsoids behave similarly in bulk fluid and are bound to the air water interface by similarly large binding energies. However, on the air-water interface their behavior is dramatically different. Anisotropic particles deform the air-water interface, producing a quadrupolar attraction that is energetically strong ($\sim10^{5}k_{B}T$).

\subsection{Particle Synthesis}

To understand how particle shape impacts particle deposition, we need particles with different shapes. We utilize micron-sized polystyrene spheres (Invitrogen), similar to the particles used in previous experiments (e.g., \cite{coffeering_nagel_nat}), and we simply modify their shape by stretching them asymmetrically to different aspect ratios \cite{ellipsoidstretch_PNAS,ellipsoid_stretch_first}. The procedures to make particles have been described previously \cite{ellipsoidstretch_PNAS,ellipsoid_stretch_first,coffee_ring_ellipsoids}, but for completeness we briefly discuss these methodologies below.

To create ellipsoidal particles, $1.3$ $\mu$m diameter polystyrene particles are suspended in a polyvinyl alcohol (PVA) gel and are heated above the polystyrene melting point ($\sim$$100$ $^{\circ}$C), but below the PVA melting point ($\sim$$180$ $^{\circ}$C) \cite{ellipsoid_stretch_first,ellipsoidstretch_PNAS}.  Polystyrene melts in the process, but the PVA gel only softens.  The PVA gel is then pulled so that the spherical cavities containing liquid polystyrene are stretched into ellipsoidal cavities.  When the PVA gel cools, polystyrene solidifies in the distorted cavities and becomes frozen into an ellipsoidal shape.  The hardened gel dissolves in water, and the PVA is removed via centrifugation. Each sample is centrifuged and washed with water at least $10$ times.

Each iteration of this process creates $\sim$$10^{9}$ ellipsoidal particles in $\sim$$50$ $\mu$l suspensions. The particles are charge-stabilized, and the resultant suspensions are surfactant-free. Snapshots of experimental particles are shown in the insets of Fig.\ \ref{droppin} b, c. The aspect ratio polydispersity is $\sim$$10\%$.  To ensure the preparation process does not affect particle deposition, our spheres undergo the same procedure, absent stretching.

Importantly, in order to ensure the PVA was not affecting our results, we performed a separate set of experiments investigating the effects of PVA on evaporating drops.  In these experiments the PVA weight percent was carefully controlled.  We found that if a sample contains more than $0.5\%$ PVA by weight, then the contact line of the drying drop depins very quickly after the drop is placed on a glass slide.  However, in samples with less than $0.5\%$ PVA by weight, the contact line behavior of the drying drop is identical to the contact line behavior in drops without PVA.  To confirm that small amounts of PVA do not affect the deposition of spheres, we added PVA ($0.45\%$ by weight) to a suspension of spheres.  During evaporation, the contact line remains pinned, and the spheres exhibit the coffee ring effect.  Further, when ellipsoids are diluted by a factor of $100$ (and thus the PVA weight percent is decreased by a factor of $100$ to an absolute maximum of $0.05\%$), the spatially uniform deposition of ellipsoids persists.

\section{Sessile Drops}

\begin{figure}
\resizebox{\columnwidth}{!}{\includegraphics{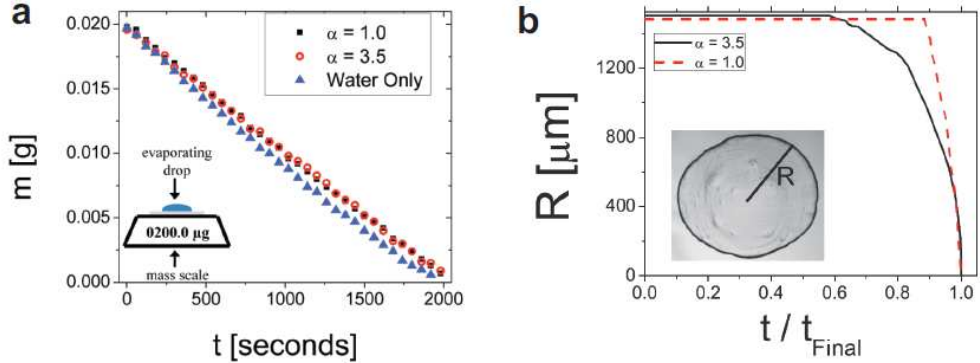}}
\caption{\label{figs1cr} a. The mass, m, of drops of different suspensions is plotted versus time, t, for evaporating drops. Suspensions of spheres ($\alpha=1.0$ black squares) and ellipsoids ($\alpha=3.5$ open red circles) are shown, as well as a drop of water absent colloids (blue triangles). Inset: Cartoon image of drop evaporating on mass balance. b. The radius, $R$, of drops of different suspensions is plotted versus time, $t$, for evaporating drops. Suspensions of spheres ($\alpha=1.0$ dashed red line) and ellipsoids ($\alpha=3.5$ black line) are shown. To facilitate comparisons, the time is normalized by the time evaporation ends ($t_{Final}$). Inset: Image defining $R$.}
\end{figure}

\subsection{Characterization of Evaporation Process}

Understanding why ellipsoids are deposited uniformly first requires that we characterize the evaporation process, i.e., we quantify the spatio-temporal evaporation profile of the suspensions. First, we are interested in the evaporation rate. To this end, we directly measure the drop mass of different suspensions ($20$ $\mu$l in volume, $6.0$ mm in radius, $\phi=0.005$) during evaporation (Fig.\ \ref{figs1cr} a). (In order to improve the accuracy of the reported evaporation rate, we utilized large-volume drops.) For all suspensions (drops of sphere suspension, drops of ellipsoid suspension, and drops of water absent colloid), the mass of each drop decreases linearly in time with very similar mass rates-of-change ($\sim10.0$ $\mu$g/s). This bulk evaporation behavior for all suspensions is consistent with steady-state vapour-diffusion-limited evaporation of spherical-cap-shaped drops with pinned contact lines \cite{drop_patterns_PRE,coffeering_nagel_nat}.\\

Next, we quantified the contact line evolution during drying, i.e., we observed when the contact line depins. Specifically, we measured the radius of the $1$ $\mu$l drops ($\phi=0.005$) during evaporation by video microscopy (Fig.\ \ref{figs1cr} b). The time at which evaporation finishes, $t_{Final}$, is clearly indicated in Fig.\ \ref{figs1cr} b as the time when the drop radius shrinks to zero.  For all samples, we observed the radius decrease by less than $10\%$ until $t=0.8\cdot t_{Final}$; i.e., the contact line remains pinned for the vast majority of the evaporation, regardless of particle shape. Note, the contact line in drops containing ellipsoids does partially depin around $t=0.7\cdot t_{Final}$; however, it does not completely depin until $t=0.8\cdot t_{Final}$.

These control experiments demonstrate that contact line behavior, capillary flow, and evaporation rates are independent of suspended particle shape. Thus, to produce qualitatively different deposits, the microscopic behaviors of individual spheres and ellipsoids in the droplets must differ.

\subsection{Particle Deposition}

\begin{figure}
\resizebox{\columnwidth}{!}{\includegraphics{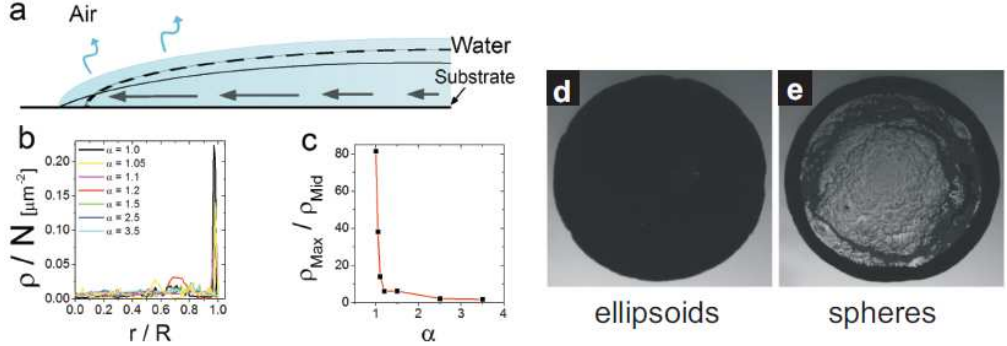}}
\caption{\label{fig1cr2} a. Schematic diagram of the evaporation process depicting capillary flow induced by pinned edges. If the contact line were free to recede, the drop profile would be preserved during evaporation (dashed line). However, the contact line remains pinned, and the contact angle decreases during evaporation (solid line). Thus, a capillary flow is induced, flowing from the center of the drop to its edges; this flow replenishes fluid at the contact line. b. Droplet-normalized particle number density, $\rho/N$, plotted as function of radial distance from center of drop for ellipsoids with various major-minor axis aspect ratios. c. The maximum local density, $\rho_{Max}$, normalized by the density in the middle of the drop, $\rho_{Mid}$, is plotted for all $\alpha$. Red lines guide the eye. d. The final distribution of ellipsoids, evaporated from a suspension with initial volume fraction $\phi=0.20$. e. The final distribution of spheres, evaporated from a suspension with initial volume fraction $\phi=0.20$.}
\end{figure}

The uniform deposition of ellipsoids after evaporation (Fig.\ \ref{droppin} b) is especially striking when compared to the heterogeneous ``coffee ring'' deposition of spheres (Fig.\ \ref{droppin} c) in the same solvent, with the same chemical composition, and experiencing the same capillary flows (Fig.\ \ref{fig1cr2} a).

To quantify the particles deposition shown in Fig.\ \ref{droppin} b and c, we determined the areal number fraction of particles deposited as a function of radial distance from the drop center (Fig.\ \ref{fig1cr2} b).  In detail, utilizing video microscopy and particle tracking algorithms, we counted the number of particles, $N_{r}$, in an area set by the annulus bounded by radial distances $r$ and $r + \delta r$ from the original drop center \cite{coffeering_nagel_nat,contactline_PRE}; here $\delta r$ is $\sim$$8$ $\mu$m.  The areal particle density $\rho(r)=N_{r}/A$, with $A =\pi((r+ \delta r)^{2}-r^{2})$. To facilitate comparisons between different samples, and eliminate small sample-to-sample particle density differences, we normalized $\rho$ by the total number of particles in the drop, N. Further, to we report $\rho(r)/N$ as a function of $r/R$, where $R$ is the drop radius, to eliminate small sample-to-sample differences in drop radii. Dilute suspensions ($\phi=0.005$) are utilized to improve image quantification. For spheres ($\alpha = 1.0$), $\rho/N$ is $\sim$$70$ times larger at $r/R \approx 1$ than in the middle of the drop. Conversely, the density profile of ellipsoidal particles is fairly uniform as a function of $r/R$ (there is a slight increase at large $r/R$). As particle shape anisotropy is increased from $\alpha = 1.0$ to $3.5$, the peak in $\rho(r)/N$ at large $r/R$ decreases. The coffee-ring effect persists for particles marginally distorted from their original spherical shape ($\alpha=1.05$ and $1.1$), but particles that are slightly more anisotropic ($\alpha=1.2$) are deposited more uniformly.

To further quantify the sharply peaked coffee-ring effect of spheres and the much more uniform deposition of the ellipsoids, we calculate and plot $\rho_{MAX} / \rho_{MID}$ (Fig.\ \ref{fig1cr2} c), where $\rho_{MAX}$ is the maximum value of $\rho$ (typically located at $r/R\approx1$) and $\rho_{MID}$ is the average value of $\rho$ in the middle of the drop ($r/R<0.25$). For spheres, $\rho_{MAX} / \rho_{MID}\approx70$.  As aspect ratio increases slightly ($\alpha=1.05$ and $1.1$) $\rho_{MAX} / \rho_{MID}$ decreases to $\sim38$ and $13$, respectively. For ellipsoids, $\rho_{MAX} / \rho_{MID}$ is more than ten times smaller than spheres. As $\alpha$ continues to increase above $1.2$, $\rho_{MAX} / \rho_{MID}$ continues to decrease, albeit at a much lower rate. Note, $\rho_{MAX} / \rho_{MID}$ was observed to be largely independent of initial volume fraction, i.e., $\rho_{MAX} / \rho_{MID}$ fluctuated by approximately $\pm$$10\%$ as volume fraction changed between $\phi = 10^{-4}$ and $0.2$.

When drops with very large packing fractions evaporate, the drop surface becomes saturated with ellipsoids. However, deposition in this limit is difficult to quantify, as at high volume fractions it is difficult to measure the local particle density.  Thus, while the particles that cannot attach to the interface are likely transported to the drop edge, it is difficult to demonstrate that this effect occurs. An experimental snapshot after evaporation of a drop of ellipsoids ($\alpha = 3.5$) initially suspended at volume fraction $\phi = 0.20$ shows that overall the coffee-ring effect is avoided, but the local density cannot be extracted (Fig.\ \ref{fig1cr2} d and e). An image of the final distribution of spheres evaporated from a suspension with initial packing fraction $\phi = 0.20$ is included for comparison.

\subsection{Real Coffee}

\begin{figure}
\resizebox{\columnwidth}{!}{\includegraphics{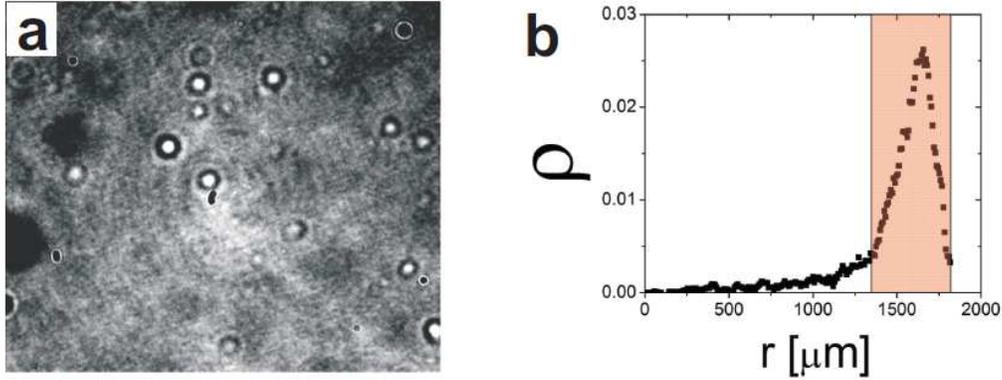}}
\caption{\label{coffeepic} a. Image of a dilute drop of coffee. Micron-sized particles in coffee appear to be relatively spherical. b. The density of adsorbed ellipsoids ($\rho$), i.e., the number of adsorbed ellipsoids per unit area, plotted versus radial position, $r$. The shaded region contains $\sim84\%$ of adsorbed particles.}
\end{figure}

Our observations thus far imply that micron-sized grains in a cup of coffee are relatively spherical. To confirm or refute this hypothesis, we prepared a microscope slide full of diluted coffee. This coffee came from the lab-building coffee machine (FilterFresh), which passes through a paper filter after a relatively short brew time ($\sim30$ seconds). While we did not ``fully'' characterize the shape of the grains we observed, qualitatively, they appeared spherical on the micron-size scale (see Fig.\ \ref{coffeepic} a). Thus, a suspension of polystyrene spheres really is an especially good model of a cup of coffee.

\begin{figure}
\caption{\label{fig2cr} a-d. Experimental snapshots at different times during the evaporation of a drop of spheres. e-h. Experimental snapshots at different times during the evaporation of a drop of ellipsoids with aspect ratio $\alpha=3.5$. i. The areal particle density, $\rho_{R}$, located within 20 \textmu m of the contact line (that is, the drop edge) as a function of time during evaporation for ellipsoidal particles.}
\resizebox{\columnwidth}{!}{\includegraphics{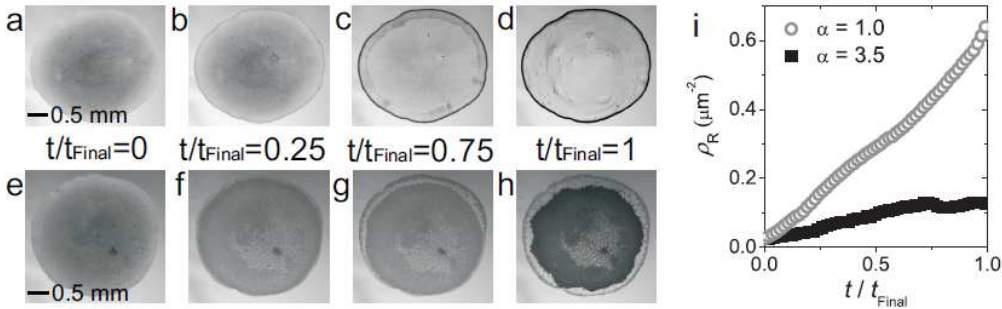}}
\end{figure}

\subsection{Interface Adsorption Profiles}

\begin{figure}
\resizebox{\columnwidth}{!}{\includegraphics{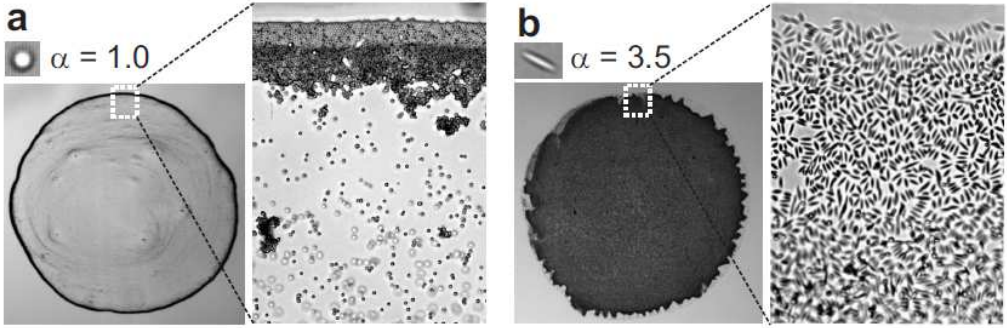}}
\caption{\label{fig3cr} Images of a region within 40 $\mu$m of the drop contact line, taken at time $t/t_{Final}=0.5$, for suspensions of spheres (a) and suspensions of ellipsoids with $\alpha=3.5$ (b). While spheres pack closely at the contact line, ellipsoids form loosely packed structures. Pictures of the entire drop after evaporation are shown and the magnified region is indicated.}
\end{figure}

The evidence suggest that the same radially outward flows are present in drops containing either spheres or ellipsoids. The deposition of spheres and ellipsoids after drying, however, is very different. In order to understand the origin of these differences better, we carried out a battery of experiments focused on the behaviors of spheres and ellipsoids on the air-water surface.

Snapshots from video microscopy show that both spheres (Fig.\ \ref{fig2cr} a-d) and ellipsoids (Fig.\ \ref{fig2cr} e-h) are carried to the drop's edges. To quantify this effect, the average areal particle density close to the contact line, $\rho_{R}=\int_{r=R-20\mu m}^{r=R}\rho(r)dr$, was measured as a function of time (Fig.\ \ref{fig2cr} i). For spheres, $\rho_{R}$ increases linearly until evaporation is complete, with a slope of $0.54$ s$^{-1}$. Conversely, the areal density of ellipsoids near the contact line stops growing at $t/t_{Final}=0.75$, and for $t/t_{Final}<0.75$, $\rho_{R}$ increases with a slope  of $0.15$ s$^{-1}$. This slope for ellipsoids is less than $1/3$ the slope for spheres, despite similar evaporation rates, capillary flows, and contact line behaviors. Thus, ellipsoid density at the drop edge grows at a slower rate than sphere density.

Next, we note that both spheres and ellipsoids strongly prefer adsorption to interface than life in the bulk drop.  Further, our experiments with ellipsoids and spheres, and previous experiments with spheres \cite{contactline_PRE,drop_patterns_PRE}, suggest that $\sim10\%$ of the particles adsorb to the air-water interface in the "central/middle" regions of the drop.  Thus most particles move toward the drop edges, and the relative drying behaviors of ellipsoids and spheres must be controlled by their behaviors near the drop edge.

To study this issue we first determine where ellipsoids adsorb on the air-water interface, i.e., we measure the number of ellipsoids that adsorb on the air-water interface as a function of radial position. The areal number density (give symbol) of ellipsoids on the air-water interface versus radial distance, (symbol) at a time immediately before the drop edge depins is given in Fig.\ \ref{coffeepic} b. The majority of particles are deposited within $\sim500$ microns of the drop's edge (at $r\sim2000$ microns). Approximately $85\%$ of the ellipsoid particles adsorb on the air-water interface in this region near the drop's edge. The properties of this interfacial region and the mechanisms by which particles attach to and move within this interfacial region play a key role in the drying process.

\subsection{Single Particle Trajectories}

What actually happens at the drop's edge? Experimental snapshots of of particles moving in the region within $40$ $\mu$m  of the drop contact line confirm that while spheres pack closely at the edge (Fig.\ \ref{fig3cr} a), ellipsoids form "loosely" packed structures (Fig.\ \ref{fig3cr} b), which prevent particles from reaching the contact line. Particles with $\alpha=1.2$ and $1.5$ pack at higher area fractions than ellipsoids with $\alpha>1.5$, resulting in larger values of $\rho_{MAX} / \rho_{MID}$ for $\alpha=1.2$ and $1.5$ and producing the small peak in $\rho(r)$ at $r/R=0.7$ for $\alpha=1.2$. The ellipsoid particle structures on the air-water interface appear to be locally arrested or jammed \cite{interface_jamming}, i.e., particles do not rearrange. Once an ellipsoid joins the collective structure, its position relative to other ellipsoids typically changes by less than $20$ nm (lower limit of our resolution), and the overall particle structure rearranges, for the most part, only when new particles attach to the interface.

Images of particles near the drop's contact line (Fig.\ \ref{fig3cr} b) reveal that unlike spheres, which are carried from the bulk all the way to the contact line (Fig.\ \ref{fig3cr} a), most ellipsoids adhere to the loosely-packed structures at the air-water interface before they reach the three-phase contact line at the drop edge. This capillary attraction has been characterized in prior experiments as long-ranged and very strong \cite{ellipsoid_fluid_interface,aniso_capil_assem,capil_force_interface,ellipsoid_wetting_yodh,capillary_interactions_ellipsoids_yodh,interface_att_furst}.

\begin{figure}
\resizebox{\columnwidth}{!}{\includegraphics{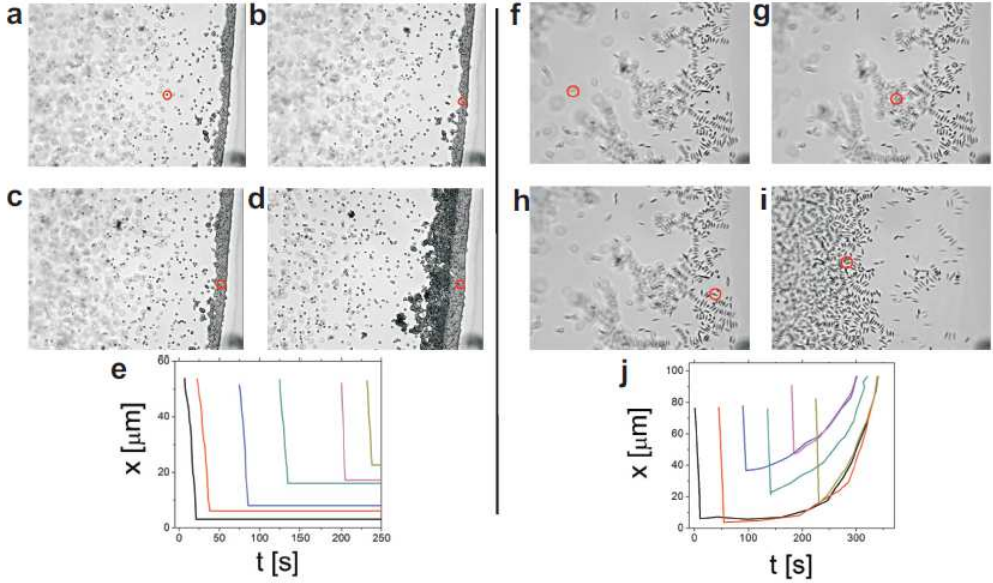}}
\caption{\label{spheretraj} a-d. Images of a drop containing spheres during evaporation at four different times ($t=1,6,26$ and $242$ seconds). The same sphere is circled in each of the four images as it travels through the bulk fluid towards the drop's edge. e. The distance from the drop's edge ($x$) for six representative spheres is plotted versus $t$. Spheres reach the drop's edge, and quickly become ``jammed,'' and cannot rearrange. f-i. Images of a drop containing ellipsoids during evaporation at four different times ($t=1,10,12$ and $622$ seconds). The same ellipsoid is circled in each of the four images. j. The distance from the drop's edge ($x$) for six representative ellipsoids is plotted versus $t$. Ellipsoids are pushed to the drop's edge through the bulk fluid, just like spheres. Once at the drop's edge, they adsorb on the air-water interface and form clusters that can migrate towards the center of the drop.}
\end{figure}

To understand the different behaviors of spheres and ellipsoids at the edge of drying drops, it is instructive to observe some individual particle trajectories. The trajectory of a single sphere is highlighted in Fig.\ \ref{spheretraj} a-d. Spheres (like the one highlighted in Fig.\ \ref{spheretraj} a-d) are pushed through the bulk fluid towards the drop's edge. When spheres reach the drop's edge, their progress is halted by a wall of spheres already at the drop's edge. Spheres then pack densely, and cannot rearrange as they jam into the ring configuration. This behavior is demonstrated quantitatively for a few typical spheres by plotting the distance ($x$) between the sphere and the drop's edge versus time (Fig.\ \ref{spheretraj} e).

Conversely, when ellipsoids reach the drop's edge, they pack loosely on the air-water interface (Fig.\ \ref{spheretraj} f-i). Notice, ellipsoids at the drop's edge do not necessarily halt the progress of other migrating ellipsoids that arrive at later times. This can be seen in Fig.\ \ref{spheretraj} f-h, as an ellipsoid approaches the drop's edge (Fig.\ \ref{spheretraj} f), passes underneath a cluster of ellipsoids on the air-water interface (Fig.\ \ref{spheretraj} g), and eventually adsorbs on the air-water interface near the drop's edge (Fig.\ \ref{spheretraj} h). As evaporation continues, ellipsoids can move along the surface of the drop towards the drop's center (Fig.\ \ref{spheretraj} i). This behavior is demonstrated quantitatively for a few typical ellipsoids by plotting $x$ versus time (Fig.\ \ref{spheretraj} j). If the air-water interface is not saturated with ellipsoids when the drop's edge depins, then the networks of ellipsoids are compressed as they are pushed towards the drop's center.

\subsection{Interface Resistance to Shear}

The loosely-packed configurations formed by ellipsoids on the interface are structurally similar to those seen in previous experiments of ellipsoids at flat air-water and water-oil interfaces \cite{ellipsoids_interface_rheology,ellipsoid_wetting_yodh,capillary_interactions_ellipsoids_yodh}. They produce a surface viscosity that is much larger than the suspension bulk viscosity, facilitating ellipsoid resistance to radially outward flows in the bulk. Note, spheres also adsorb onto the interface during evaporation.  However, spheres do not strongly deform the interface \cite{ellipsoid_wetting_yodh}, and they experience a much weaker interparticle attraction than ellipsoids \cite{interface_att_furst}; therefore, the radially outward fluid flows in the bulk and interface easily push spheres to the drop's edge \cite{contactline_PRE}.

In order to quantify the ability of interfacial aggregates of ellipsoids to resist bulk flow, we calculated the Boussinesq number, B$_{0}$, for ellipsoids with $\alpha=3.5$. Specifically, B$_{0}$ is the ratio of the surface drag to the bulk drag: B$_{0}=\frac{G'}{\tau L}$ where $\tau$ is shear stress from bulk flow, G' is the elastic modulus of the interfacial layer, and $L$ is the probed lengthscale \cite{boussinesq_number}. B$_{0}$ varies spatially with the average areal particle density on the air-water interface. Here, we calculate B$_{0}$ in a region within $40$ $\mu$m of the pinned contact line.

\begin{figure}
\resizebox{\columnwidth}{!}{\includegraphics{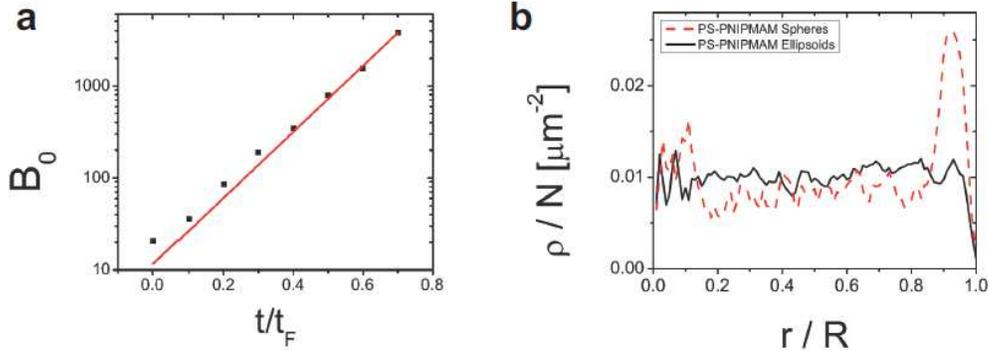}}
\caption{\label{figs3cr} a. The Boussinesq number, B$_{0}$, for ellipsoids with $\alpha=3.5$ is plotted versus time, t, normalized by the time evaporation finishes, t$_{F}$. The red line is the best exponential fit. b. Droplet-normalized particle number density, $\rho/N$, plotted as function of radial distance (normalized by the drop radius) from center of drop for core-shell polystyrene-PNIPMAM spheres (red dashed line) and core-shell polystyrene-PNIPMAM ellipsoids (solid black line). The hydrophillic PNIPMAM coating does not qualitatively affect the deposition of spheres and ellipsoids.}
\end{figure}

We first calculated B$_{0}$ at an early time (t $= 0.1$ t$_{F}$).  The shear stress can be calculated from the particle velocity and drop height via $\tau \approx \mu v / L$, where $\tau$ is shear stress, $\mu$ is viscosity, and $L$ is the drop height. At an early time (t $= 0.1$ t$_{F}$) $\tau\approx3\cdot10^{-4}$ Pa. About $40\%$ of the surface is covered with ellipsoids.  Previous experimental studies measured the shear modulus, $G'$, of the interfacial monolayer as a function of surface coverage area fraction \cite{ellipsoids_interface_rheology}. We measured the surface coverage area fraction in our system as a function of time. This measurement enabled us to utilize the values of $G'$ reported in \cite{ellipsoids_interface_rheology} ($G' \approx10\cdot^{-3}$ N/m).  The probed lengthscale, $L$, is at most $0.01$ m (i.e., the drop diameter). Thus, at t $=0.1$t$_{F}$, B$_{0}\sim300$.  This calculation is performed at different times during evaporation, until the final stage of evaporation when the aggregate of ellipsoids begins flowing towards the drop center (Fig.\ \ref{figs3cr} a). We found that $\tau$ grows linearly with particle velocity, which we observe to increase by a factor of $\sim2$ during evaporation. However, $G'$ grows exponentially with the ellipsoidal area fraction\cite{ellipsoids_interface_rheology}, and area fraction increases by a factor of $\sim3$.  Thus, the exponential growth of $G'$ dominates this calculation, and B$_{0}$ grows exponentially with time: B$_{0}\propto exp(\frac{t}{0.12 t_{F}})$. Finally, note that for spheres, B$_{0}<1$.  Thus, the measured dimensionless Boussinesq number clearly demonstrates that clusters of ellipsoids on the air-water interface can resist shear from radially outward fluid flows, and make sense of the fact that these clusters are not pushed to the drop's edge. Conversely, clusters of spheres on the air-water interface cannot resist shear and are pushed along the air-water interface to the drop's edge where the join the coffee-ring deposit.

\begin{figure}
\resizebox{\columnwidth}{!}{\includegraphics{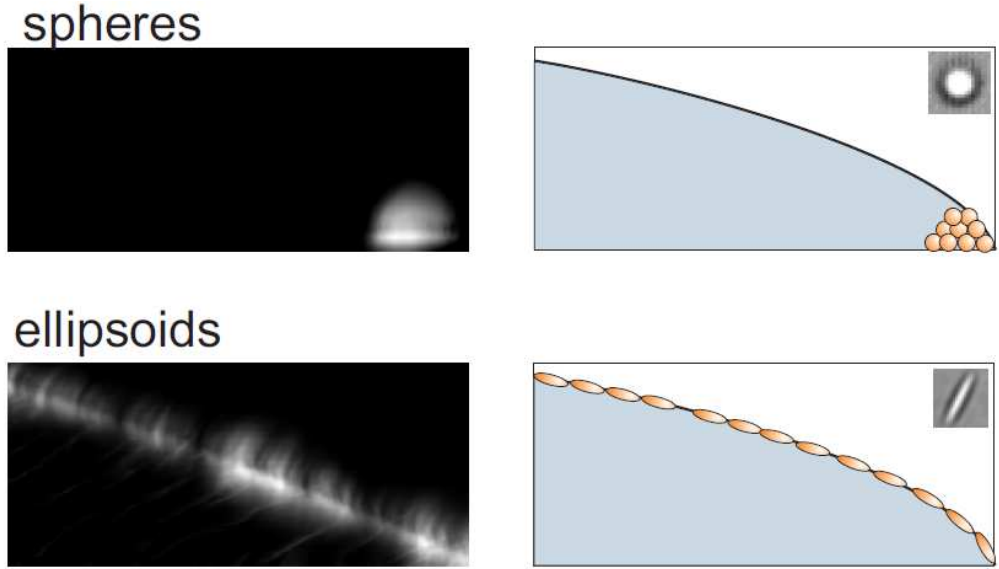}}
\caption{\label{figconfocal} Confocal projections of suspension of spheres (top) and ellipsoids with $\alpha=2.5$ (bottom) onto the z-r plane in cylindrical coordinates. While spheres are efficiently transported to the contact line, ellipsoids sit at the air-water interface.}
\end{figure}

\subsection{Confocal Microscopy}

We have already shown that ellipsoids sit (largely without moving) at the air-water interface. Here we utilize confocal microscopy to directly measure the location of ellipsoids (and spheres) during evaporation. Confocal snapshots are shown in Fig.\ \ref{figconfocal}. By integrating the brightness of each pixel over a period of $0.05$ seconds, only particles that are roughly stationary during this time period appear in the images.  Snapshots are then projected onto a side-view of the drop. The confocal images confirm that ellipsoids sit at the air-water interface (Fig.\ \ref{figconfocal} bottom), while spheres are carried all the way to the contact line (Fig.\ \ref{figconfocal} top).

\subsection{Other Anisotropic Particles and Hydrophobicity Issues}

In order to assess the generality of this effect, we have analyzed three additional types of anisotropic particles.  One parameter potentially important for this process is particle hydrophobicity. Hydrophilic particles, for example, are perhaps less likely to adsorb onto the air-water interface than hydrophobic particles and might equilibrate differently on the interface than hydrophobic particles; thus the hydrophilic ellipsoids could have different deposition during drying. To investigate the effect of hydrophilicity, we obtained suspensions of spherical and ellipsoidal polystyrene-PNIPMAM core-shell particles, i.e., polystyrene particles coated with PNIPMAM. We evaporated these suspensions at $23$ $^{\circ}$C; at this temperature, PNIPMAM is hydrophilic. The core-shell hydrophilic spheres exhibit the coffee ring effect (Fig.\ \ref{figs3cr} b). Conversely, despite their hydrophilicity, core-shell ellipsoids are deposited uniformly. In fact, these core-shell ellipsoids form the same type of loosely-packed ellipsoid networks on the drop surface as polystyrene ellipsoids absent PNIPMAM (Fig.\ \ref{figs3cr} b).

Further, we have evaporated suspensions of actin filaments and Pf$1$ viruses.  In each of these suspensions, the contact line depins at very early times. To prevent this early depinning, we add a small amount of $50$ nm diameter fluorescent polystyrene spheres ($\sim1\%$ by weight); these spheres help to pin the contact line until the final stage of evaporation (t $>$ $0.8$ t$_{F}$) via self-pinning \cite{drop_patterns_PRE}. The spheres in each suspension exhibit the coffee-ring effect. Both the actin filaments and Pf$1$ viruses in suspension, however, are deposited relatively uniformly.  (Note, the mean major axis length for Pf$1$ viruses is $\sim2$ $\mu$m; the mean minor axis length for Pf$1$ viruses is $\sim6$nm. The mean major axis length for actin filaments is $\sim20$ $\mu$m; the mean minor axis length for actin filaments is $\sim20$nm.\\)

\begin{figure}
\resizebox{\columnwidth}{!}{\includegraphics{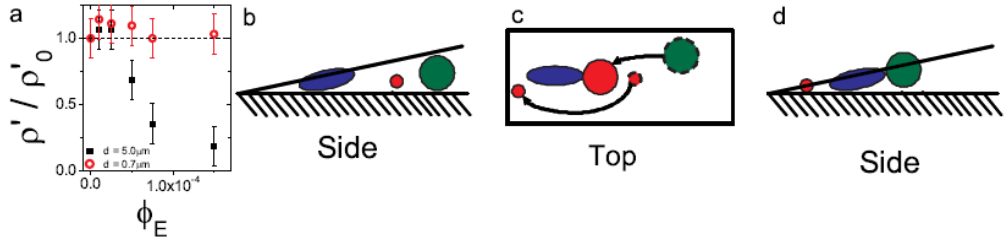}}
\caption{\label{fig4cr} a. The deposition of mixtures of spheres and ellipsoids are characterized by the ratio $\rho'=\rho_{Max}/\rho_{Mid}$, where $\rho_{Max}$ is the maximum local density and $\rho_{Mid}$ is the density in the middle of the drop, as a function of ellipsoid volume fraction, $\phi_{E}$.  Two sizes of particles are studied: $d=5.0$ $\mu$m (black squares), $d=0.7$ $\mu$m (red circles), where $d$ is the particle diameter. To best capture the evolution of the deposition as $\phi_{E}$ increases, $\rho'$ is normalized by $\rho'_{0}$, the value of $\rho'$ when there are no ellipsoids present, i.e., $\phi_{E}=0$.  The coffee ring effect persists for mixtures of small spheres and ellipsoids, but the coffee ring is destroyed for mixtures of large spheres and ellipsoids. Error bars represent the statistical uncertainty that results from finite bin sizes. b-d. Cartoon depicting capillary flow that carries suspensions of spheres and ellipsoids to the drop's edge. The left panel is a side view at an early time, the right panel is a side view at a later time, and the center panel is a top view showing particle trajectories in between those times. Spheres that are smaller than the ellipsoid continue to travel all the way to the edge, and exhibit the coffee ring effect. Spheres larger than the ellipsoids are affected by deformations of the air-water interface, and join the ellipsoids in loosely packed structures forming at the interface.}
\end{figure}

\subsection{Mixtures of Spheres and Ellipsoids}

Lastly, we investigate the effects of mixing ellipsoids and spheres. A small number of ellipsoids were added to suspensions of different sized spheres. We then evaporate drops of suspensions containing both ellipsoids and spheres.  Our initial hope was that a small number of ellipsoids could dramatically change the deposition behavior of spheres in suspension.

To simplify this study, we concentrated on two different aspect ratios: spheres ($\alpha=1.0$) and ellipsoids ($\alpha=3.5$). The ellipsoids were stretched from particles of diameter $d=1.3$ $\mu$m; each suspension contains spheres suspended at a volume fraction $\phi=0.02$.  Evaporative deposits are characterized as a function of ellipsoid volume fraction $\phi_{E}$ via $\rho'(\phi_{E})=\rho_{Max}/\rho_{Mid}$ (Fig.\ \ref{fig4cr} a).

Suspensions containing smaller spheres with $d=0.7$ $\mu$m along with the ellipsoids at volume fractions ranging from $\phi_{E}=0$ to $1.5\times10^{-4}$ were evaporated.  The coffee-ring effect persists for these small spheres, regardless of how many ellipsoids are added to the initial suspension (Fig.\ \ref{fig4cr} a). Small spheres can easily navigate under or through the loosely packed ellipsoid networks, and thus reach the drop's edge (Fig.\ \ref{fig4cr} b-d).

For comparison,  we evaporated suspensions containing larger spheres with $d=5.0$ $\mu$m, along with the same ellipsoids at the same volume fractions utilized previously. For small ellipsoid volume fraction ($\phi_{E}\leq2.5\times10^{-5}$), the evaporating suspensions still exhibit the coffee-ring effect. However, for larger $\phi_{E}$, the coffee ring is diminished; for sufficiently large $\phi_{E}$, i.e., $\phi_{E}\approx1.5\times10^{-4}$, the coffee-ring effect is avoided (Fig.\ \ref{fig4cr} a). Larger spheres adsorb onto the air-water interface farther from the drop edge than do the smaller ellipsoids. Absent ellipsoids, spherical particles form closely-packed aggregates. In the presence of ellipsoids, the spheres instead become entangled in the loosely-packed ellipsoid networks, thus eliminating the coffee ring effect (Fig.\ \ref{fig4cr} b-d). Therefore, large spherical particles can be deposited uniformly simply by adding ellipsoids.

\subsection{Sessile Drop Future Directions}

The ability to deposit particles uniformly is desirable in many applications \cite{printing_convec_assemb}.  Unfortunately, most proposed methods for avoiding the coffee ring effect require long multistage processes, which can be costly in manufacturing or require use of organic solvents which are sometimes flammable and toxic (e.g. \cite{marangoni_reverses_coffeering,monolayer_jaeger}).  Here we have shown that by exploiting a particle's shape, a uniform deposit can be easily derived from an evaporating aqueous solution. The results presented here further suggest that other methods of inducing strong capillary interactions, e.g., surface roughness \cite{roughness_cap_attract}, may also produce uniform deposits.

Additionally, open questions about the behavior of ellipsoids in drying drops persist. Specifically, one may have thought the drop's edge would quickly saturate with ellipsoids during evaporation, and ellipsoids subsequently arriving would then be deposited in a coffee-ring stain. However, ellipsoids (and their collective structures) clearly migrate towards the drop's center during evaporation, in the process creating room for more ellipsoids to adsorb on the air-water interface near the drop's edge. It is unclear why ellipsoids move towards the drop's center. Inward fluid flows along the drop's surface push networks of ellipsoids towards the drop's center, thus making room for more ellipsoids to adsorb on the air-water interface. Alternatively, the energetic interactions of the ellipsoids on the air-water interface may play an important role in this inward migration. However, a complete understanding of this inward motion has been elusive and will require more experimental and theoretical investigation.

\section{Evaporation and Deposition from Confined Colloidal Drops}

\begin{figure}
\resizebox{\columnwidth}{!}{\includegraphics{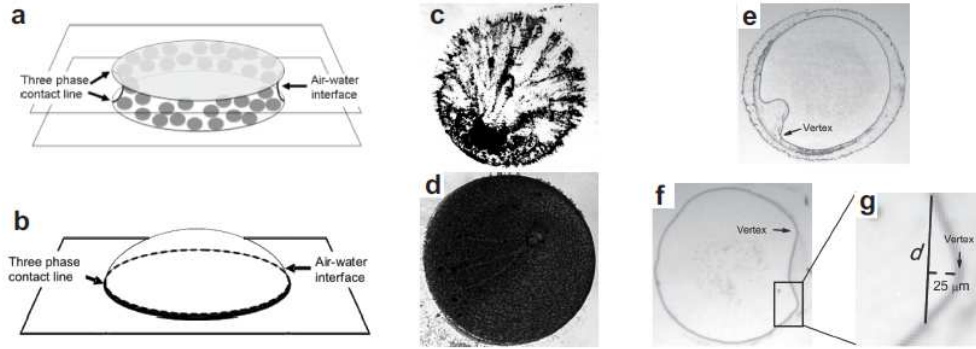}}
\caption{\label{confined} Cartoon depicting droplet evaporating in a confined geometry (a) and an open geometry (b). The particle-populated air-water interface and three phase contact lines are labeled. c,d. Image of the final deposition of particles with major-minor diameter aspect ratio $\alpha=1.0, 3.5$ (c,d, respectively). e,f. Sample images (top-view) of buckling events for confined drops containing anisotropic particles with $\alpha=1.2$ and $1.5$ (e,f, respectively). g. Rim width, $d$ (solid line), is defined here in a magnified image of a buckled region, as the interface full-width $25$ $\mu$ m from the vertex of the bent air-water interface (see dashed line).}
\end{figure}

The mechanism that produces a uniform coating from particles suspended in drying sessile drops requires the presence of an air-water interface that spans the entire area covered by the drop. A drop confined between two glass plates is a completely different beast. In this case, the air-water interface is only present at the drop edges Fig.\ \ref{confined}a and Fig.\ \ref{figs1ce} c (a sessile drop is shown for comparison in Fig.\ \ref{confined}b). Thus, the mechanisms that produce uniform coatings in "open" or sessile drops are unlikely to be present in confined drops. To illustrate these spectacular differences, in Fig.\ \ref{confined} c,d, we again see that suspended particle shape produces dramatically different depositions.  The confined drops don't even exhibit the conventional coffee ring effect.  Rather, spheres and slightly stretched spheres are deposited heterogeneously, and anisotropic ellipsoids are distributed relatively more uniformly.  In this section, we show how one can understand these deposition effects.  Important clues are revealed through consideration of the mechanical properties of the air-water interfaces, and changes thereof as a result of adsorbed particles.

Recent experiments have explored evaporation of confined drops containing spheres \cite{confined_evap,evap_2d_preview,confined_evap_theory,meniscus_lithography}, and their behaviors differ dramatically from sessile drops containing spheres. In the confined case, particles are pushed to the ribbon-like air-fluid interface, and, as evaporation proceeds, the particle-covered air-water interface often deforms and crumples (Fig.\ \ref{confined} e and f). The buckling behaviors exhibited by these ribbon-like colloidal monolayer membranes (CMMs) in confined geometries are strongly dependent on the geometric shape of the adsorbed particles, and the buckling events appear similar to those observed in spherical-shell elastic membranes \cite{landau_elasticity,crumpling_sphere_gompper}. Before buckling events occur, particles are densely packed near the three-phase contact line, regardless of particle shape. Further, because the particle volume fraction in the drop is relatively low, these membranes essentially contain a monolayer of particles, i.e., buckling events occur before multilayer-particle membranes form.

These experiments utilize the same micron-sized polystyrene ellipsoids described above in Section 2.2 and in \cite{ellipoid_stretch_firstdye,ellipsoidstretch_PNAS,ellipsoid_stretch_first}. Drops of suspension are confined between two glass slides separated by $38.1$ $\mu$m spacers (Fisher Scientific) and allowed to evaporate; qualitatively similar results are found for chambers made from slightly hydrophobic cover slips. We primarily study the drops with initial particle volume fraction $\phi=0.01$. (Qualitatively similar results are found for volume fractions ranging from $\phi = 10^{-4}$ to $0.05$.) The confinement chambers are placed within an optical microscope wherein evaporation is observed at video rates at a variety of different magnifications. This approach also enables measurement of the surface coverage, i.e., the fraction of the air-water interface coated with particles, prior to buckling events. We find that for spheres and ellipsoids the surface coverage areal packing fraction is $\sim0.70 \pm 0.05$.

\subsection{Theory of Buckled Quasi-2D Membranes}

To understand this buckling phenomenon, the elastic properties of the air-water interface with adsorbed particles, i.e., the elastic properties of the CMMs, must be quantified. To this end, the analytical descriptions of elastic membranes are extended to our quasi-2D geometry wherein observations about bending and buckling geometry are unambiguous. This theoretical extension has been described previously (see \cite{confined_evap_ellipsoids} and its associated supplemental online material), but for completeness and clarity of presentation we discuss it more completely below.

\begin{figure}
\resizebox{\columnwidth}{!}{\includegraphics{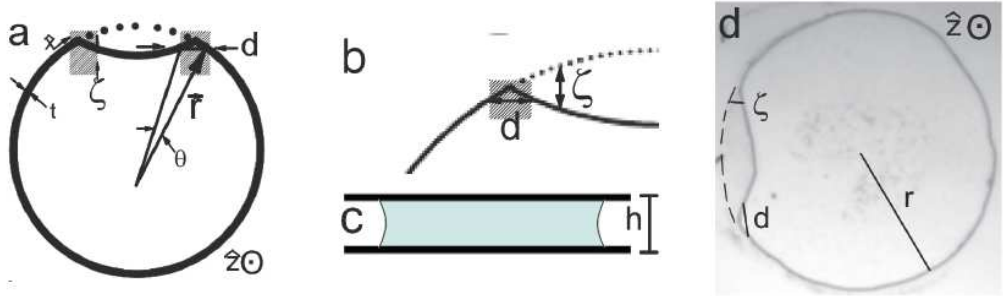}}
\caption{\label{figs1ce} a. Buckling event cartoon defining rim full-width, $d$, drop radius, $r$, interface displacement, $\zeta$, membrane thickness, $t$, in-plane direction along membrane surface, $\hat{x}$, angle, $\theta$, and out-of-plane direction, $\hat{z}$. The dotted line represents the initial membrane configuration (before the buckling event). The regions containing all buckling and stretching energy are shaded. All un-shaded regions are unstretched and unbent. b. Magnified buckling event cartoon defining rim full-width, $d$, and interface displacement, $\zeta$. The regions containing all buckling and stretching energy are shaded. c. Side view cartoon defining chamber height, $h$. d. Example of buckling event for a confined drop containing anisotropic particles with $\alpha=1.5$. The rim width, $d$ (solid line), drop radius, $r$, interface displacement, $\zeta$, and out-of-plane direction, $\hat{z}$, are defined here. Dashed line indicates initial position of membrane. }
\end{figure}

Following the same procedure as \cite{landau_elasticity}, we first describe the stretching and bending energy associated with membrane buckling events. Membrane stretching energy can be written as $E_{S}=0.5\int E u^{2}dV$, where $E_{S}$ is the total membrane stretching energy, $E$ is the 2D Young's modulus, $u$ is the strain, and the integrand is integrated over the membrane volume. For a thin, linearly elastic material, $u$ does not change much in the direction perpendicular to the surface, so $E_{S}\cong 0.5\int Eu^{2}dA$, where the integral is calculated over the membrane surface area. The unstretched region has $u=0$.  Further, even in the stretched/buckled membrane, most of the deflected region has $u=0$, since its configuration is identical to the undeflected membrane except that its curvature is inverted (Fig.\ \ref{figs1ce} a,b).  Thus, the only region under strain is the ``rim'' of the deformation (Fig.\ \ref{figs1ce} a,b).  If the entire membrane had experienced a constant radial displacement of $\zeta$, its radius would change from $r$ to $r+\zeta$, and the circumference would change from $2\pi r$ to $2\pi (r+\zeta)$.  Then the membrane strain would be $u=2\pi \zeta / 2\pi r = \zeta / r$.  On the other hand, if (as is the case for our samples) the displacement is confined to a small region subtended by angle $\theta$, then the in-plane length of this region changes from $\theta r$ to $\theta (r + \zeta)$, and the total strain in the membrane is $u = \theta \zeta / \theta r = \zeta / r$.  Again, this estimate assumes that the interfacial deflection does not change in the $z$-direction (out-of-plane), i.e., $\partial\zeta/\partial z\approx0$. Within these approximations, $E_{S} \cong 0.5\int E(\zeta / r )^{2}dA$.  The integral is readily performed over an area normal to the glass plates described by $A\approx d h$, where $d$ is the in-plane length of the deflected region, and $h$ is the chamber height.  Thus, $E_{S}\approx 0.5 E(\zeta/r)^{2}dh$.

The membrane bending energy can be written as $E_{B} = 0.5\int \kappa K_{C}^{2} dA $, where $E_{B}$ is the total bending energy, $\kappa$ is the bending rigidity, and $K_{C}$ is the membrane curvature. Here, the curvature is $K_{C} \approx \partial ^{2} r(\theta) / \partial x^{2}$, where $x$ is the coordinate in-plane along the membrane (see Fig.\ \ref{figs1ce} a,b). The first derivative can be written as $\partial r(\theta) / \partial x \approx \zeta / d$, as $\zeta$ is the change in the membrane position over a distance of approximately $d$ in the $x$ direction. The second derivative can then be estimated as $\partial^{2} r(\theta) / \partial x^{2} \approx \zeta / d^{2}$, as the first derivative changes from $0$ in the undeflected region to $\zeta / d$ in the deflected region of approximate length $d$. Therefore, $K_{C}\approx \zeta / d^{2}$. (This approach again assumes that the second derivative of the deflection in the $z$-direction is small, i.e., $\partial^{2}\zeta/\partial z^{2}\approx0$.) The integral is readily performed over an area described by $A\approx d h$, and $E_{B}\approx 0.5\kappa h \zeta^{2} / d^{3} $.

The total energy from the deflection is $E_{TOT}=E_{S}+E_{B}=0.5E (\zeta/r)^{2}dh+0.5\kappa h \zeta^{2} / d^{3}$. This energy is concentrated within the deflected rim (i.e., with width $d$).   Membranes will buckle in the way that minimizes their energy.  To derive this condition, we minimize the total deflection energy with respect to $d$, i.e., $\partial E_{TOT} / \partial d = E (\zeta/r)^{2}h-3\kappa h \zeta^{2} / d^{4} =0$. Minimizing the total bending and stretching energy gives the relation, $\kappa / E = d^{4} / (3r^{2})$. Thus, by measuring $d$ and $r$ in a series of drops with the same particles and membrane characteristics, we can experimentally determine $\kappa / E$. (Interestingly, $\zeta$ drops out of the calculation, i.e., a precise determination of $\zeta$ is not necessary for this calculation within the assumptions listed above.  Also, note that this calculation is independent of the depth of the invagination; the only requirement is that the deflection minimizes total membrane energy. Finally, note that this derivation assumes that the interfacial displacement varies little in the $z$-direction, i.e., the air-water interface deflects the same distance at the top, middle, and bottom of the chamber.)

In practice we measure $d$ as the rim full-width located $25$ $\mu$m from the rim vertex (see Fig.\ \ref{confined} g and Fig.\ \ref{figs1ce} a, b and d). The exact value of $d$, however, is not very sensitive to measurement protocol. For example, defining $d$ as the full-width at $20$ $\mu$m or $30$ $\mu$m from the rim vertex changes $d$ by approximately $20$ percent.

\begin{figure}
\resizebox{\columnwidth}{!}{\includegraphics{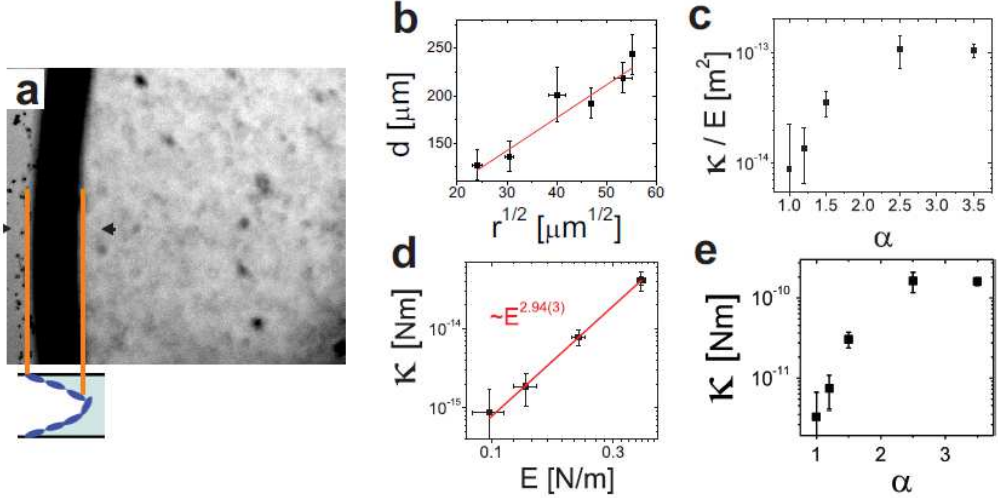}}
\caption{\label{figs2ce} a. Experimental image of air-water interface demonstrating how the radius of curvature is measured. Red lines represent the inner and outer edges of the air-water interface, as shown in the cartoon below. b. $d$ is plotted versus the square root of the drop radius, $r$. c. Ratio of the bending rigidity, $\kappa$, to the Young's modulus, $E$, is plotted versus $\alpha$. d. $\kappa$ versus $E$, where $E$ comes from previously reported measurements and calculations. The line represents the best power law fit. e. $\kappa$ versus $\alpha$.}
\end{figure}

This simple experimental approach enables us to extract the ratio of CMM bending rigidity, $\kappa$, to its Young's modulus, $E$, from measurements of $d$ and $r$ across a series of drops from the relation $\kappa/E=d^{4}/(3r^{2})$.  With all other parameters constant, e.g., particle anisotropy, particle surface coverage, etc., this formula predicts that $d\propto \sqrt{r}$. In Fig.\ \ref{figs2ce} b we show results from evaporated drops of particles with anisotropy $\alpha=1.2$ and with different initial values of $r$, plotting $d$ versus $\sqrt{r}$.  A good linear relationship is observed (coefficient of determination, R$^{2}=0.93$), implying that our analysis is self-consistent. Similar high quality linear results were found for other values of $\alpha$.

In principle, the air-water interface can be distorted in the $z$-direction as well as in-plane. The analysis thus far has assumed these distortions are small, and it is possible to check that these corrections are small. Using bright field microscopy, we can identify the inner and outer position of the air-water interface and thus estimate the radius of curvature in the $z$-direction \cite{evap_2d_preview} (Fig.\ \ref{figs2ce} a). We find that the radius of curvature is approximately equal to the chamber thickness ($\sim38.3 \mu$m $\pm 1 \mu$m) both before and after buckling events. The relevant partial derivatives are then $\partial\zeta/\partial z \leq 1/38.3$ and $\partial\zeta/\partial z \leq 1/(38.3^{2})$; therefore the corrections to theory are indeed small.

\subsection{Dependence of Bending Rigidity on Particle Shape}

We extract and plot $\kappa/E$ for evaporating drops of particles with different $\alpha$ (Fig.\ \ref{figs2ce} c). Notice, $\kappa/E$ increases with increasing $\alpha$, implying that as particle shape becomes more anisotropic, $\kappa$ increases faster than $E$, i.e., $\kappa/E$ is larger for ellipsoids ($\alpha=2.5$ and $3.5$) than for spheres ($\alpha=1.0$).

Since we measure the ratio $\kappa/E$, in order to isolate the bending rigidity we require knowledge of the Young's modulus of the membrane. Previous experiments have observed that the CMM Young's modulus increases with $\alpha$ \cite{ ellipsoid_monolayers_bulkmodulus, ellipsoids_interface_rheology,ellipsoid_mono_rheo_theory,youngs_modulus_spheres,rheology_2D_gel}. For particles with $\alpha=1.0$ and $2.5$, we use previously reported values of the bulk modulus \cite{ellipsoid_monolayers_bulkmodulus}, $B$, the shear modulus \cite{ellipsoids_interface_rheology}, $G'$, and the relationship $E=4BG' / (B+G')$ in order to extract the CMM Young's modulus. We were unable to find data for $\alpha=1.2,1.5,$ or $3.5$, so we linearly interpolated from reported values of $B$ and $G'$. Using these previously reported values, we obtained $E=0.098, 0.14, 0.22, 0.39$ and $0.39$ N/m for $\alpha=1.0,1.2,1.5,2.5$ and $3.5$, respectively.

\begin{figure}[ht]
\begin{minipage}[b]{0.47\linewidth}
\centering
\includegraphics[width=\textwidth]{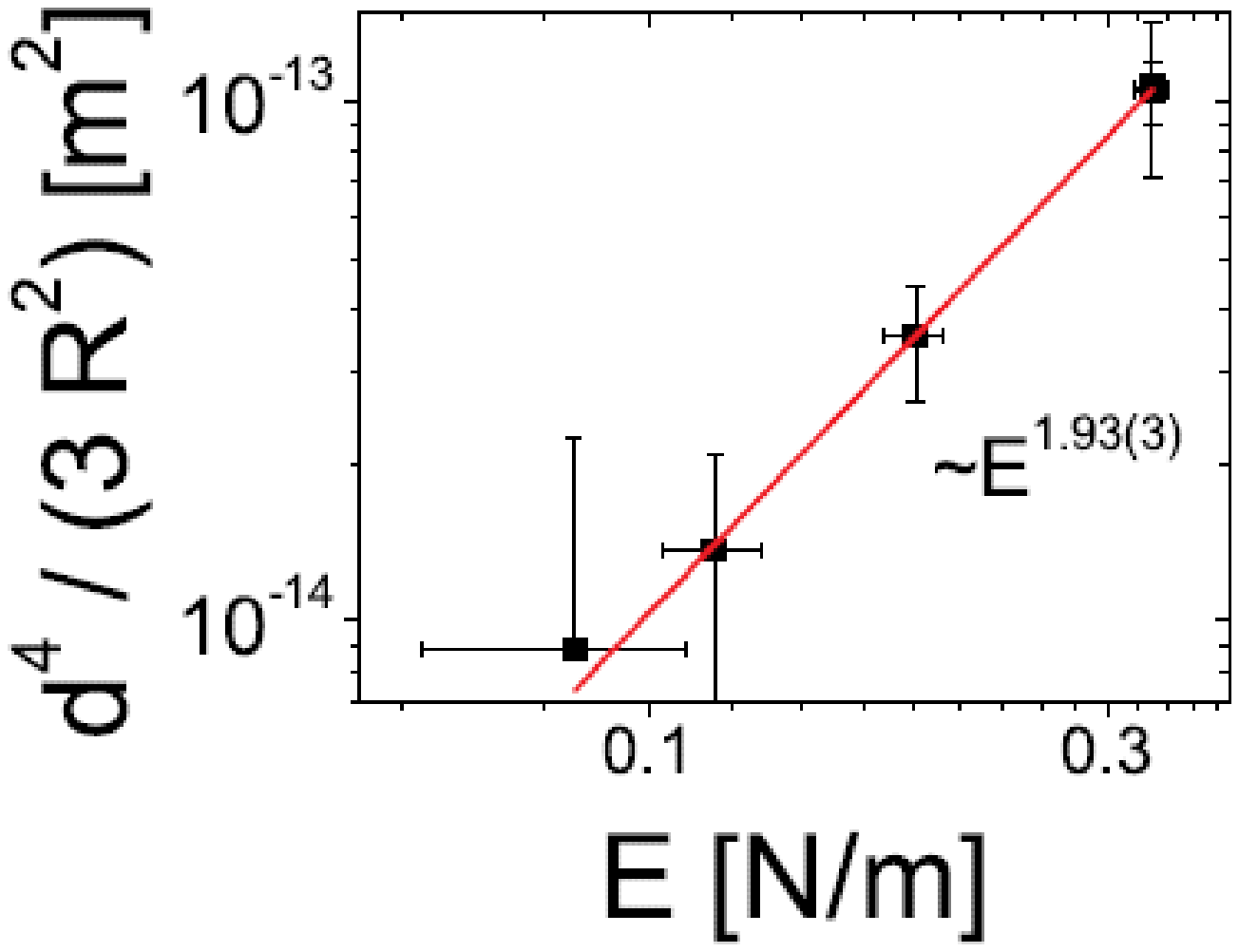}
\caption{As a consistency check, $d^{4}/(3r^{2})$ is plotted versus $E$. The line represents the best power law fit.}
\label{figs3ce}
\end{minipage}
\hspace{0.45cm}
\begin{minipage}[b]{0.47\linewidth}
\centering
\includegraphics[width=\textwidth]{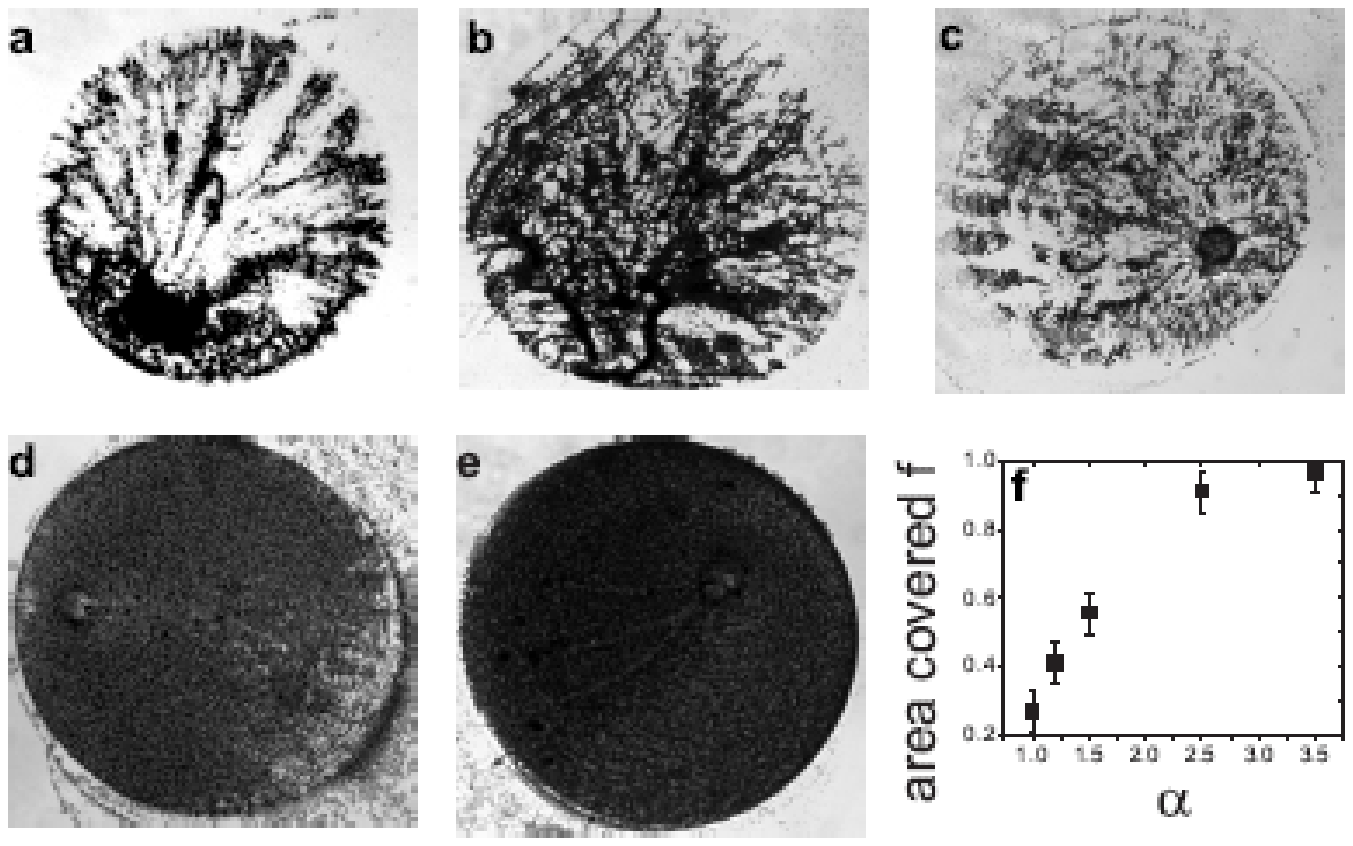}
\caption{Image of the final deposition of particles with major-minor diameter aspect ratio $\alpha=1.0,1.2,1.5,2.5,3.5$ (a-e, respectively). f. The area fraction covered by particles after evaporation is complete, \textit{f}, for suspensions of particles as a function of their aspect ratio $\alpha$.}
\label{fig2ce}
\end{minipage}
\end{figure}

Utilizing these previously reported measurements and calculations of $E$ we are able to plot $\kappa$ versus $E$ (Fig.\ \ref{figs2ce} d). The best power-law fit finds that $\kappa \propto E^{2.94(3)}$. Interestingly, this observation is consistent with theoretical models which predict $\kappa \propto E^{3}$ \cite{landau_elasticity}. However, the full physical origin of this connection is unclear. Further, while at first glance it may seem contradictory to claim that $\kappa/E=d^{4}/(3r^{2})$ and $\kappa \propto E^{3}$, these formulae are consistent. A simple elastic model assumes that $E=Yt$ and $\kappa=Yt^{3}$, where $Y$ is the 3D Young's modulus and $t$ is the membrane thickness \cite{landau_elasticity}. Based on this model, $\kappa=E^{3} / Y^{2}$, so $\kappa/E=E^{2} / Y^{2}$. Thus, $\kappa/E=E^{2} / Y^{2}=d^{4}/(3r^{2})$.  To test this prediction, we plot $d^{4}/(3r^{2})$ versus $E$ (Fig.\ \ref{figs3ce}). The best power law fit is $d^{4}/(3r^{2}) \propto E^{1.92(3)}$, implying that these two seemingly contradictory equations are in fact consistent. Note, this simple elastic model suggests that $Y \approx 19$ kPa for all $\alpha$, which is similar to stiff jello. Finally, our estimates of CMM bending rigidity are given in (Fig.\ \ref{figs2ce} e). Clearly, membrane bending becomes much more energetically costly with increasing particle shape anisotropy.

\subsection{Particle Deposition in Confined Geometries}

Finally, we turn our attention to the problem we initially hoped to understand: the consequences of increased bending rigidity on particle deposition during evaporation processes in confined geometries. As should be evident from our discussion in Sections 1-3, substantial effort has now yielded understanding of the so-called coffee-ring effect and some ability to control particle deposition from sessile drops \cite{coffeering_nagel_nat,convect_assemb_nagayama_nature,evaporation_theory,monolayer_jaeger,marangoni_reverses_coffeering,coffeering_reverse_capforce,surfactant_rev_CReffect,marangoni_benard_PRL,disjoiningpressure_assemb,coffee_ring_electrowetting,sommer,coffee_ring_ellipsoids}.  Much less is known, however, about particle deposition in confined geometries, despite the fact that many real systems \cite{evap_model_soil,confined_evap_liquids,wetsand} and applications \cite{evap_applications,porous_microfluidics} feature evaporation in geometries wherein the air-water interface is present only at the system edges. Recent experiments have explored evaporation of confined drops containing spheres \cite{confined_evap,evap_2d_preview,confined_evap_theory,meniscus_lithography}, and their behaviors differ dramatically from sessile drops containing spheres. In the confined case, as noted previously, particles are pushed to the ribbon-like air-fluid interface, and, as evaporation proceeds, the particle-covered air-water interface often undergoes the buckling events which we have quantified in Sections 4.1 and 4.2.

We find that deposition depends dramatically on suspended particle shape.  The final deposition of particles is shown for $\alpha=1.0,1.2,1.5,2.5,3.5$, in Fig.\ \ref{fig2ce} a-e, respectively. Spheres and slightly stretched spheres are deposited unevenly, while anisotropic ellipsoids are distributed much more homogeneously.

To quantitatively describe the final deposition of particles, we plot the fraction of initial droplet area covered by deposited particles after evaporations, \textit{f} (as introduced in \cite{drop_patterns_PRE}), as a function of particle anisotropy $\alpha$ (Fig.\ \ref{fig2ce} f). Specifically, we divide the area into a grid of ($8$ $\mu m$ X $8$ $\mu m$) squares. A region is considered to be covered if its area fraction within the square is greater than $0.36$. (Note, for uniformly deposited particles, the area fraction (based on the initial volume fraction, initial volume, chamber height, and particle size) would be $\sim$$0.4$. Thus, the threshold we utilize is $\sim90\%$ of this uniformly deposited area fraction). The number of covered regions is then normalized by the total number of squares in the grid, thus producing \textit{f}. The fraction of area covered with particles is observed to increase with $\alpha$. For $\alpha=1.2$ and $1.5$, \textit{f} increases modestly. For $\alpha=2.5$, the deposition is very uniform, and for $\alpha=3.5$, virtually the entire area is covered uniformly.

\begin{figure}
\resizebox{\columnwidth}{!}{\includegraphics{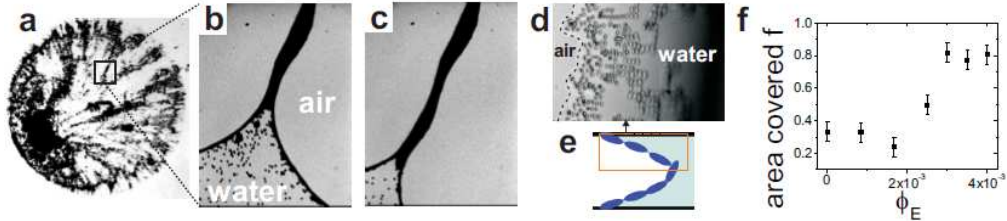}}
\caption{\label{fig3ce} a. Image of the final deposition of particles with major-minor diameter aspect ratio $\alpha=1.0$. The box indicates the deposit left behind by the event depicted in (b) and (c). b. Image of a pinned region of the air-water interface ($\alpha=1.0$).  When the pinned section does not ``snap'' off, it leaves behind a channel. c. At a later time ($\sim$$100$ seconds after (c)), the channel extends, and more particles flow into it, producing a very heterogeneous deposition. d. Image of a colloidal monolayer near the three phase contact line in a drop containing ellipsoids ($\alpha=3.5$). The three phase contact line is labeled with a dashed line on the left side of the image. Particles are adsorbed on the air-water interface, forming a monolayer, as evidenced by the fact that particles become more out of focus, from left to right, as the air-water interface curves. A cartoon below shows a side view of the experimental image (e). f. The fraction of area covered by particles, \textit{f}, for suspensions of $200$ nm diameter spheres doped with different amounts of ellipsoids, represented by the ellipsoid volume fraction, $\phi_{E}$.}
\end{figure}

The mechanisms that produce the uneven deposition of spheres and slightly stretched particles and the uniform deposition of ellipsoids are revealed by high magnification images (Fig.\ \ref{fig3ce} a-e). Colloidal particles locally pin the contact line and thereby locally prevent its motion. So-called self-pinning of the air-water interface can occur even in very dilute suspensions, i.e., $\phi<10^{-4}$ \cite{drop_patterns_PRE}. As evaporation continues in suspensions of spheres or slightly anisotropic particles, the CMM interface bends around the pinning site (Fig.\ \ref{fig3ce} a-c).  Then, either it pinches off, leaving particles behind, or it remains connected to the pinned site, leading to fluid flow into the narrow channel that has formed. The latter flow carries particles towards the pinning site (Fig.\ \ref{fig3ce} b and c), thus producing streaks of deposited particles (see Fig.\ \ref{fig3ce} a-c). Temporal and spatial variations along the interface due to these described effects lead to heterogeneous deposition of spherical particles during evaporation.

Conversely, when ellipsoids adsorb onto the air-water interface (forming ribbon-like CMMs, see Fig.\ \ref{fig3ce} d), they create an elastic membrane with a high bending rigidity. The bending rigidity of ellipsoid-populated CMMs can be approximately two-orders of magnitude larger than sphere-populated CMMs (see Fig.\ \ref{figs2ce}).  Thus, while ellipsoids may also pin the contact line, bending of the CMM interface around a pinned contact line is energetically costly. Microscopically, bending requires the energetically expensive rearrangement of ellipsoids aggregated on the CMM; attractive particle-particle capillary interactions on the air-water interface must be overcome for bending, even at very small $\phi$. Conversely, bending a sphere coated CMM costs relatively little energy, as sphere-sphere capillary interactions on the interface are relatively weak \cite{capillary_interactions_ellipsoids_yodh,ellipsoid_wetting_yodh,interface_att_furst}. Thus, as the confined drop continues to evaporate, the ellipsoid coated CMM does not bend. It recedes radially, depositing ellipsoids near the contact line during this drying process.

\subsection{Mixtures of Spheres and Ellipsoids}

As we already demonstrated mixing spheres and ellipsoids in sessile drops presents qualitatively new scenarios. It is natural to investigate the deposition of mixtures of spheres and ellipsoids in confined geometries. To this end suspensions of $200$ nm spheres ($\alpha$$=$$1.0$) with $\phi$$=$$0.02$ were combined with suspensions containing micron-sized ellipsoids ($\alpha$$=$$3.5$) at lower volume fractions, $\phi$$=0$ to $4.0\times10^{-3}$. The resulting colloidal drops were evaporated in the same confined geometries already utilized. The addition of a very small number of ellipsoids has no effect on the deposition of spheres ($\phi \leq 1.7\times10^{-3}$). However, the addition of a larger, but still small number of ellipsoids produces a uniform deposition of both ellipsoids and spheres, i.e., $f\approx0.8$, despite the fact that spheres significantly outnumber ellipsoids ($10^{3}$-$10^{4}$) (Fig.\ \ref{fig3ce} e).

Again, the high bending modulus produced by ellipsoids on the CMM helps explain the observations.  Both spheres and ellipsoids attach to the air-water interface. Ellipsoids deform the air-water interface, creating an effective elastic membrane with a high bending rigidity.  When enough ellipsoids are present, pinning and bending the interface becomes energetically costly and the spheres (and ellipsoids) are deposited as the interface recedes.

Further, this behavior in confined geometries is different than that of sessile drops (see Section 3.8 and \cite{coffee_ring_ellipsoids}). From this perspective, it is somewhat surprising that small spheres are deposited uniformly from droplets doped with small numbers of ellipsoids and confined between glass plates.

Interestingly, this method of producing a uniform deposition is similar to convective assembly techniques wherein the substrate, or a blade over the substrate, is pulled away from the contact line in a colloidal suspension; a thin film is thus formed that leads to the creation of a monolayer of particles (e.g., \cite{convective_assembly_blade, ellipsoids_furst, convect_assemb_nagayama_nature, convect_assemb_xtalcoating, nagayama_langmuir, convect_assemb_dip, convect_assemb_crystal, convect_assemb_coffeering, convect_capp_assemb, convect_assemb_ellipsoid}). Unlike many other convective assembly techniques, the present experimental system has neither moving nor mechanical parts. Uniform coatings are created essentially as a result of shape-induced capillary attractions which produce CMMs that are hard to bend.

\subsection{Evaporation of Drops in Confined Geometries: Summary}

Colloidal drops evaporating in confined geometries behave quite differently the evaporating sessile drops.  Ellipsoids adsorbed on the air-water interface create an effective elastic membrane, and, as particle anisotropy aspect ratio increases, the membrane's bending rigidity increases faster than its Young modulus. As a result, when a drop of a colloidal suspension evaporates in a confined geometry, the different interfacial elastic properties produce particle depositions that are highly dependent on particle shape. The ability to increase CMM bending rigidity by increasing particle shape anisotropy holds potentially important consequences for applications of CMMs. For example, increased bending rigidity may help stabilize interfaces (e.g., Pickering emulsions \cite{emulsions_ellipsoids_softmatter}) and thus could be useful for many industrial applications, e.g., food processing \cite{food_emulsions,coffee_ring_ellipsoids_newsviews}. In a different vein, the observations presented here suggest the buckling behavior of CMMs in confined geometries may be a convenient model system to investigate buckling processes relevant for other systems, e.g., polymeric membranes \cite{elastic_membrane_polymer}, biological membranes \cite{elastic_membrane_bio_cells}, and nanoparticle membranes \cite{elastic_membrane_nanoparticles}.

\section{Surfactant Effects on Particle Deposition from Drying Colloidal Drops}

In the previous sections, we showed how particle shape influences the behaviors of drying drops containing colloidal particles.
For sessile drops we found that particle anisotropy could be employed to overcome the coffee ring effect; for drops in confinement, we found that particle anisotropy dramatically affected the bending rigidity of the air-water interfaces which in turn modified particle deposition during drying. Besides particle shape, many other ideas have been observed, developed, and utilized over the years to manipulate the drying behaviors of colloidal drops \cite{monolayer_jaeger,marangoni_reverses_coffeering,coffeering_reverse_capforce,surfactant_rev_CReffect,marangoni_benard_PRL,disjoiningpressure_assemb}.
In the final section of this review paper, we describe our foray into the effects of added surfactants.

A surfactant is a surface-active molecule that consists of a hydrophobic and a hydrophilic part. In water, surfactant molecules populate the air-water interface with their hydrophobic parts ``sticking out of the water,'' thereby reducing the water's surface tension (which is paramount for the cleaning effects of soaps or dish washers).
In an immiscible mixture of water and oil, surfactants populate the interfaces between components, thus stabilizing the emulsion. In an evaporating drop of an aqueous colloidal suspension, surfactants give rise to other effects.

Herein we describe video microscopy experiments which investigate how a small ionic surfactant (mostly) affects particle deposition in drying drops; these surfactants induce a concentration-driven Marangoni flow on the air-water interface and a strong ``eddy''-like flow in the bulk that prevents particles from depositing in the coffee ring and thus suppresses the coffee ring effect for spheres.

\begin{figure}
\resizebox{0.5\columnwidth}{!}{\includegraphics{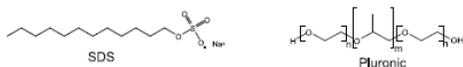}}
\caption{\label{fig:structures} Chemical structures of SDS and Pluronic surfactant.}
\end{figure}

Although we focus here primarily on small ionic surfactants, we have explored the effects of a variety of surfactants.  In general, common types are small, ionic surfactants, e.g., sodium dodecyl sulfate (SDS), or large, polymeric ones, e.g., pluronics; the chemical structures of both examples are shown in Fig.~\ref{fig:structures}.
Accordingly, surfactants can affect deposition phenomena in a variety of ways. For example, it was found that SDS can change the deposition patterns from aqueous colloidal drops \cite{deegan00}.
In different experiments, surfactant is sprayed onto the drop \cite{nguyen02,truskett03}, leading to complex patterns as a result of thermodynamic transitions between different phases formed by the surfactant.

If the surface tension is heterogeneous on a liquid surface (e.g., the air-water interface of a drop) a flow is induced from regions of low to high surface tension.  This effect is the so-called Marangoni flow. Such Marangoni flows can result from different temperatures at drop edge and center, e.g., because of different evaporation rates and slow diffusive heat transfer; thus, in principle such a flow should be present at the air-water interfaces of drying liquid drops \cite{zhang82,savino02,hu05b,ristenpart07}. Indeed, Marangoni radial flows towards the center of a drop have been found in small drops of octane \cite{hu06}. In water, however, such temperature-dependent Marangoni flows are suppressed \cite{deegan00,hu06,savino02,hu05b}.

In addition to temperature-driven changes of the surface tension, surfactant-driven Marangoni flows have been suggested to explain the relatively uniform deposition of dissolved polymer from droplets of organic solvent containing surfactant \cite{kajiya09}.
When the local concentration of surfactant molecules at the pinned contact line increases due to the coffee-ring effect, then the surface tension of the drop decreases locally, and a gradient in surface tension arises. This gradient has been suggested as the source of continuous Marangoni flow towards the center of the drop \cite{kajiya09}.

Herein, we first investigate the mechanism of a small ionic surfactant, SDS, on the evaporation of aqueous colloidal systems and their resulting particle coatings \cite{still12}.
The experiments demonstrate that such small ionic surfactants do indeed produce Marangoni flows in colloidal droplets, not only in agreement with the model suggested for polymer solutions \cite{kajiya09}, but also providing a first direct visualization.
We further demonstrate how the ``Marangoni eddy'' can lead to uniform particle deposition during drying, thereby undermining the coffee ring effect.

At the end of this section on surfactants we show preliminary experiments which demonstrate that large polymeric surfactants like Pluronic F-127 influence the evaporation of drops in a strikingly different way than small ionic surfactants.
In this case, contact line pinning is prevented, leading to a uniform particle deposition.
We suggest an explanation of this behavior as due mainly to an increase of viscosity near the contact line, which is a result of high polymer concentration because the dissolved polymeric surfactant is transported to the contact line by the coffee-ring flow.

\subsection{Experimental Methods}\label{experimental}

The procedure for these experiments has been described previously \cite{marangoni_eddies}, but for completeness and presentation clarity we briefly discuss these methodologies below. We focus on a few representative systems of evaporating drops with the small ionic surfactant SDS or the large polymeric surfactant Pluronic F-127, respectively, and we attempt to elucidate rules governing their behavior.

We employed aqueous suspensions of colloidal polystyrene (PS) particles (diameter $d=1330$~nm, synthesized by surfactant free radical emulsion polymerization, and stabilized by sulfate groups \cite{goodwin74}).
Suspensions were prepared with deionized water, filtered by a millipore column, and then the suspensions of PS spheres and SDS (Sigma-Aldrich) or Pluronic F-127 (BASF) (in different compositions) were thoroughly mixed by a vortexer and ultrasonicated for five minutes.

Evaporation experiments were observed using a brightfield microscope with air objectives (magnification 5x to 100x).
Clean hydrophilic glass substrates (Fisher Scientific) were used as evaporation substrates.
(Note: Qualitatively similar results were found on hydrophobic cover slips.)
The drop volume was about 0.05~\textmu L, leading to deposition coatings with diameters of 1 to 3~mm.
The evaporation process was recorded by a video microscopy (camera resolution 658x494 pixel, 60 frames per second) with total evaporation times between 2 and 4 minutes.
All experiments were repeated several times in order to identify a consistent concentration-dependent behavior.
Photographs of the entire deposit, obtained after evaporation, were taken by combining up to four high-resolution photographs when the deposition area was larger than the microscope field of view \cite{still12}.

\subsection{Surfactant Driven Particle Deposition and Marangoni Eddies}

\begin{figure}
\resizebox{\columnwidth}{!}{\includegraphics{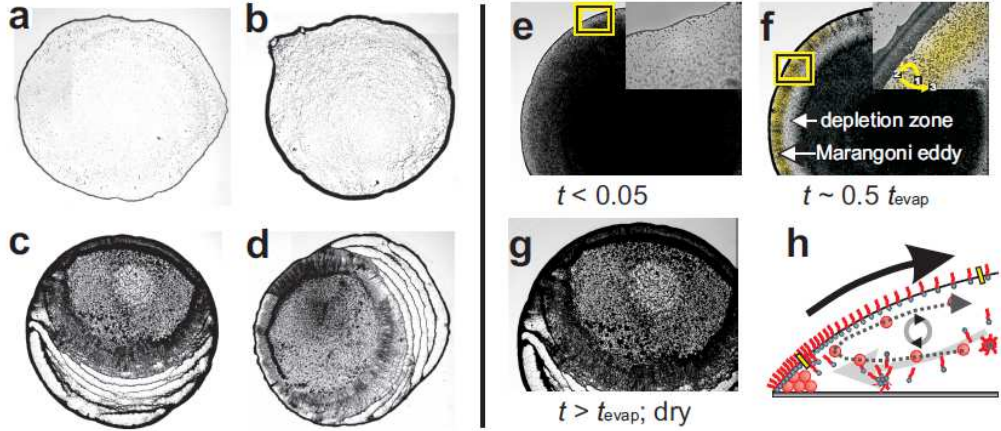}}
\caption{\label{fig:SDS1} \label{fig:sds2}Deposition patterns formed by completely evaporated water drops containing 0.5 wt\% PS particles ($d$=1330 nm) and different concentrations of SDS (\textbf{a} no SDS, \textbf{b} 0.05 wt\%, \textbf{c} 0.5 wt\%, and \textbf{d} 1.0 wt\%) on hydrophilic microscope slides. (cf. Fig. 1 in \cite{still12}) Evaporation process of a water drop containing 0.5 wt\% PS particles ($d$=1330~nm) and 0.5 wt\% SDS at different states of the drop evaporation ($t_{evap}$: total evaporation time).
	\textbf{e} $t<0.05t_{evap}$; Initial coffee ring like motion. In the inset at higher magnification, it can be seen that the ring consist of only a few particles.
	\textbf{f} $t\approx 0.5t_{evap}$; ``Marangoni eddy'' (highlighted yellow): surfactant concentration driven flow as described in the text and depicted in (h).
	In the inset, an exemplary single sphere's motion is highlighted by the numbers 1-3, respectively, at the time of the picture, 0.25~s later (i.e., sphere next to the edge), and 0.5~s later (i.e., sphere repelled).
	\textbf{g} $t>t_{evap}$ (dry), cf. (c).
	\textbf{h} Cartoon describing the ``Marangoni eddy'': SDS molecules from the bulk are pushed to the pinned contact line, where they concentrate at the water/air interface, leading to a locally decreased surface tension and, thus, a surface Marangoni flow towards the center of the drop, where it is balanced by the outward-directed coffee ring flow. (cf. Figs. 2 and 3 in \cite{marangoni_eddies}}
\end{figure}

Fig. \ref{fig:SDS1}a-d shows top views of the deposition pattern of an aqueous suspension of PS spheres (0.5 wt\%) (\textbf{a}) and similar suspensions but with different concentrations of SDS ranging from 0.05 wt\% to 1.0 wt\% (\textbf{b-d}).
The coffee-ring effect is observed in sample \textbf{a}, i.e., the vast majority of spheres are deposited in a thin ring located at the initial pinned contact line, and very few particles are deposited in the center of the drop. The deposition changes slightly upon adding a small amount of SDS (0.05 wt\%, \textbf{b}). Specifically, the coffee-ring broadens and more particles are deposited in the center of the drop.

At higher SDS concentrations (0.5 wt\% (\textbf{c}) and 1.0 wt\% (\textbf{d})), however, the deposition pattern changes drastically.
Instead of a single ring at the initial pinned contact line, tree-ring like structures are observed with several distinct deposition lines.
These tree-ring deposition structures can be explained by stick-slip dynamics of the drop's contact line along a large part of the perimeter; after the edge depins and the drop shrinks, a new contact line stabilizes very quickly via self-pinning by other PS particles in suspension \cite{deegan00}. Mutiple depinning and repinning events produce the observed pattern. Inside these tree ring-like structures, both systems exhibit relatively uniform depositions of spheres about their centers, surrounded by dark ``flares''.

Clearly, the addition of surfactant has large influence on how the particles are deposited.
But how can the effects we observe be explained?
To answer this question, we studied the temporal evolution of drops by video microscopy during evaporation.
Fig. \ref{fig:sds2} shows snapshots of the drying drop in Fig. \ref{fig:SDS1}c (0.5 wt\% SDS) at three different stages: \textbf{d} at the begining of evaporation, \textbf{e} after about 50\% of the total evaporation duration, $t_{evap}$, has passed, and \textbf{f} after evaporation is complete. High magnification images taken from a similar drop with identical composition are shown as well.

The first thing we noticed is that when the evaporation starts, drop behavior appears identical to that of drops without SDS, i.e., the contact line is pinned and spheres initially pack densely at the drop's edge arrange in a densely packed structure at the edge (see Fig. \ref{fig:sds2}e inset). The image gets progressively darker towards the drop center where the drop is thickest as particles in the bulk are evenly distributed.

However, even at this early stage of evaporation, some spheres flow towards the drop edge but do not reach it.
Rather, as they approach the edge, they are repelled back towards the drop center.
As we know from the early studies of the coffee ring effect \cite{deegan00}, with advancing evaporation time, the flow towards the edge becomes stronger. In our experiment, more and more particles approach the edge but do not reach it. These particles appear to be captured within a certain region of the drop, which is highlighted yellow in Fig. \ref{fig:sds2}f. They form a broad corona, i.e., an outer rim distinctly different and separated from the inner part of the drop, located between the relatively uniform dark center and the coffee-ring.

We describe the dark part of the corona as a ``Marangoni eddy'' or circulating region of PS spheres that are transported towards and away from the drop edge throughout the drying process (see Fig. \ref{fig:sds2}f). PS spheres are pulled into the eddy (see Fig. \ref{fig:sds2}f), leading to a locally reduced number of particles in the depletion zone (that explains why this region is less dark then the other regions). The trajectory of an individual sphere is marked by the three numbers (1, 2, 3) in Fig. \ref{fig:sds2}f. Initially, the sphere is approximately in the middle of the eddy (1). After $\sim0.25$~seconds, the sphere is pushed radially outward, i.e., towards the coffee ring (2). However, after another$\sim0.25$~seconds, the sphere is pushed radially inward, i.e., towards the region between the eddy and the depletion zone (3).
Video microscopy shows us that the same behavior is observed for virtually all of the particles at later times during the evaporation (see supporting online material for \cite{marangoni_eddies})

\subsection{Discussion of Marangoni Eddies}

The experiments provide evidence that the observed deposition behavior is dominated by a surfactant-driven Marangoni effect. As noted above, related surfactant-driven phenomena were recently observed in drying polymer solutions containing oligomeric fluorine-based surfactants \cite{kajiya09}. However, the previously studied polymer solutions differ qualitatively from the aqueous colloidal suspensions presented here. Drying polymer solutions can exhibit gelation \cite{okuzono09}. Further, the local surface tension in drying drops of polymer solutions depends on the local solute concentration \cite{poulard07,kajiya09b}.

Observing particle motion in real-time facilitates a comprehensive understanding of this phenomenon. A cartoon of the mechanism is shown in \ref{fig:sds2}g. The ``eddy'' forms in between the yellow bars in Fig. \ref{fig:sds2}g, which corresponds to the highlighted region in Fig. \ref{fig:sds2}f in (top view). Shortly after a drop is created, some surfactant molecules (pictured in the cartoon as hydrophilic ``heads'' with hydrophobic ``tails'') adsorb on the water/air interface. Note, the air-water interface is the energetically preferred location for the amphiphilic SDS molecules. However, electrostatic repulsion of anionic heads prevents them from forming maximally dense steric equilibrium packing. Additionally, at sufficiently high concentration SDS molecules are also dissolved in the bulk, either freely or as micelles (see Fig. \ref{fig:sds2}d).

As was the case with no surfactant is added, the contact line is initially pinned. Thus, the outward convective flow that is responsible for the coffee-ring effect is present and transports spheres and SDS molecules to the drop edge.
As a result, the air-water interface near the contact line becomes more concentrated with SDS molecules which locally decrease the interfacial surface tension $\gamma$. This creates a surface tension gradient along the air-water interface, $\nabla\gamma$, which is resolved by a Marangoni flow from low to high $\gamma$. This strong surface flow penetrates into the bulk fluid, so that it can carry colloidal spheres near, but not on the interface, towards the drop center.

As spheres flow towards the drop center, the local SDS concentration decreases (and the local $\gamma$ increases), and the Marangoni flow weakens.
Eventually, the radially outward bulk convective flow that drives the coffee-ring effect dominates the radially inward Marangoni flow. Particles that travel to this point are carried towards the drop's edge once again; the process then repeats. Although the SDS molecules are too small to be observed optically, SDS molecules likely participate in the eddy as well. Otherwise, the surface would become saturated with SDS, which, in turn, would end or at least weaken the Marangoni flow. Thus, particles are trapped in a circulating flow driven by the local surfactant-concentration, which we call the ``Marangoni eddy''. We find that the Marangoni flows become stronger at SDS concentrations above its critical micelle concentration (cmc) in water (8.3~mM$\approx$0.2~wt\%\cite{cifuentes97}).

Again, as was the case for drops without SDS, at late times the contact line is observed to depin. However, due to the presence of the Marangoni eddy, when the final depinning occurs, many particles are left in the more central regions of the bulk, because the eddy prevented them from attaching to the edge. After the contact line depins, the particles that remain in the bulk are deposited onto the substrate as the radially-inward-moving contact line passes them. The radially-inward contact line thus leaves behind a relatively uniform particle deposition in the drop center. The formation of a ``Marangoni eddy'' is a prerequisite for the relatively uniform particle deposition in the drop center; it prevents many particles from depositing at the drop's edge, and thereby delays their deposition until times when the contact line has depinned.

\section{Other Surfactants}

\begin{figure}
	\centering
		\includegraphics[width=0.90\textwidth]{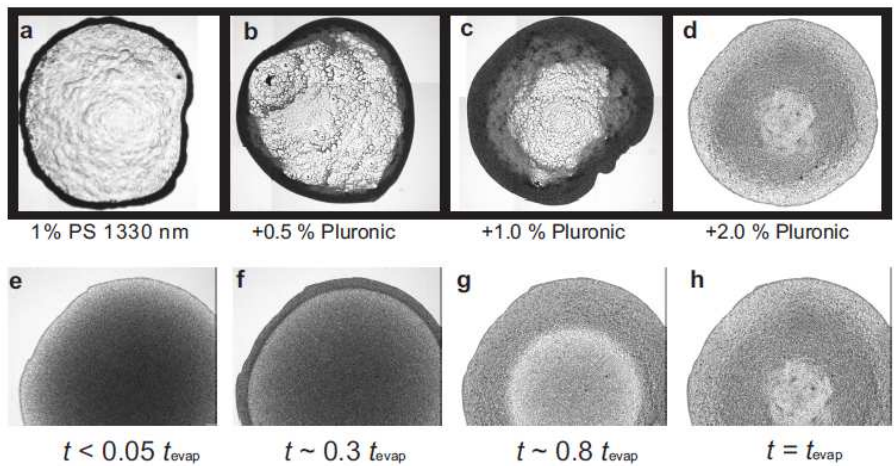}
	\caption{\textbf{a-d} Set of depositions from dried water drops containing 1.0 wt\% PS particles ($d$=1330 nm) and different concentrations of Pluronic F-127 (indicated under the pictures). The drops were evaporated on a hydrophilic microscope slide, the initial drop volume was about 0.05~\textmu l, and the resulting deposition area is between 1-2~mm in diameter.
	\textbf{e-h} Three sets of snapshots of evaporating water drops containing 1.0 wt\% PS particles ($d$=1330 nm) and 2 wt\% Pluronic F-127 at different states of the evaporation.}
	\label{fig:pluronic1}
\end{figure}

Lastly, we describe preliminary experiments with non-ionic triblock polymer surfactants such as Pluronic F-127 is present ($n\approx106$, $m\approx70$, cf. Fig.~\ref{fig:structures}).
Fig.~\ref{fig:pluronic1} is analogous to Fig.~\ref{fig:SDS1}, but with 1~wt\% PS particles and 0.5, 1.0, and 2.0 wt\% Pluronic. . Pluronic F-127 is a relatively large molecule ($M_w\approx12600$~g/mol); in the investigated drops, the amount of Pluronic is about the same (by weight) as colloidal spheres.

The addition of Pluronic leads to a systematic change in the deposition pattern; as more surfactant is added, the initial ring becomes broader, and eventually the entire area is coated uniformly (on a macroscopic scale) with PS spheres. Additionally, at lower concentrations of Pluronic, complex deposition patterns appear in the center of the drop area. Video microscopy indicates that a Marangoni eddy is not present during drying.
Instead, Pluronic induces an early depinning of the contact line and a loose packing of spheres

Why doesn't a surfactant like Pluronic F-127 produce a ``Marangoni eddy?''
Like SDS, the Pluronic F-127 molecule is amphiphilic and the coffee ring effect should also transport it to the edge were it could, in principle, give rise to the same flow effects as SDS.

Cui et al. attribute similar behaviors found in samples containing poly(ethylene oxide) to a combination of several effects \cite{cui12}. Their most important argument is that dissolved polymer is transported to the drop edge where it leads to a dramatic increase of viscosity such that the suspended colloidal particles are immobilized before they reach the contact line. This argument does not provide insight about why the contact line should move, but we speculate that the hydrophobicity of the deposited polymer (in our case, surfactant) may play a crucial role.

\begin{figure}
\resizebox{\columnwidth}{!}{\includegraphics{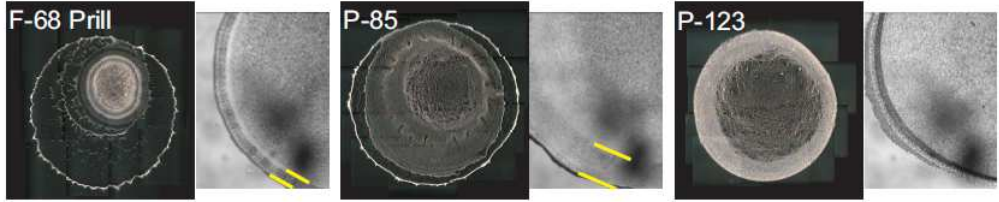}}
\caption{\label{fig:pluronic6} Combination of a bright field microscope picture at $t\approx 0.5 t_{evap}$ (\textit{left}) and a dark field picture of the dry residue after complete evaporation (right) for three different pluronics (each 0.5~wt\% PS particles (1330~nm) and 1~wt\% surfactant); (left) Pluronic F-68 Prill ($n\approx80$, $m\approx30$, $M\approx8.4$~kDa), (middle) Pluronic P-85 ($n\approx26$, $m\approx40$, $M\approx4.6$~kDa), (right) Pluronic P-123 ($n\approx20$, $m\approx70$, $M\approx5.8$~kDa). For \textbf{a} and \textbf{b}, where a Marangoni effect is observed, the width of the Marangoni waltz is indicated by short yellow lines.}
\end{figure}

Interestingly, some other surfactants in the Pluronic family, principally the same structure (cf. Fig.~\ref{fig:structures} but with different block lengths $n,m$, show a strong Marangoni eddy. Specifically, different Pluronics (all BASF) were explored and their deposition patterns are shown in Fig.~\ref{fig:pluronic6}; these surfactants include Pluronic F-68 Prill ($n\approx80$, $m\approx30$, $M\approx8.4$~kDa), P-85 ($n\approx26$, $m\approx40$, $M\approx4.6$~kDa), and P-123 ($n\approx20$, $m\approx70$, $M\approx5.8$~kDa). For all three Pluronics, Fig.~\ref{fig:pluronic6} shows a snapshot of the evaporating drop at $t\approx0.5t_{evap}$ on the left and a dark-field microscopy photograph of the deposition after drying is completed.
Interestingly, for F-68 (a) and P-85 (b), a Marangoni eddy similar to that seen for SDS appears. Correspondingly, the deposition pattern of drops with these surfactants is more similar to the case of SDS than to Pluronic F-127.

On the other hand, Pluronic P-123 (c) leads to the same phenomenon as F-127, revealing a mostly uniform, loose deposition with no evidence of a Marangoni eddy.
A comparison of the molecular properties of all poloxamer surfactants shows that neither the total molecular weight nor the ratio $m/n$ is the parameter that governs the drop evaporation.

In total, the examples in this review illustrate the complexity of the coffee ring problem.
In a few carefully controlled situations is the deposition of particles from a drying drop is dominated by a single effect. In most cases, several cooperative or antagonistic effects act at the same time, preventing easy predictions and phenomenological understanding of the underlying principles needed to open new pathways towards further technological application of the coffee ring effect or its circumvention \cite{han12}.

\acknowledgments
We thank Matthew Gratale, Matthew A. Lohr, Kate Stebe, and Tom C. Lubensky for helpful discussions. We also gratefully acknowledge financial support from the National Science Foundation through DMR-0804881, the PENN MRSEC DMR11-20901, and NASA NNX08AO0G.

\bibliographystyle{varenna}


\end{document}